\documentclass[aps,showpacs,prb,twocolumn,superscriptaddress,floatfix]{revtex4-2}
\usepackage{graphicx,bm,amssymb,amsmath,dcolumn,hyperref}
\usepackage{multirow}
\usepackage[shortlabels]{enumitem}
\usepackage{color}
\usepackage{subfigure}  
\usepackage{epstopdf}
\usepackage{bbm}
\usepackage{graphicx}
\usepackage{hyperref}
\usepackage{threeparttable}
\usepackage{amsthm}
\usepackage{mathtools}
\usepackage{color}
\usepackage{tikz}
\usepackage{float}
\usepackage{empheq}
\usepackage{algorithm}
\usepackage{algpseudocode}
\usepackage{array}

\begin{document}
	
\title{Fate of Berezinskii-Kosterlitz-Thouless Paired Phase in Coupled $XY$ Models}

\author{Tianning Xiao}
\affiliation{Hefei National Research Center for Physical Sciences at the Microscale and School of Physical Sciences, University of Science and Technology of China, Hefei 230026, China}

\author{Youjin Deng}
\email{yjdeng@ustc.edu.cn}
\affiliation{Hefei National Research Center for Physical Sciences at the Microscale and School of Physical Sciences, University of Science and Technology of China, Hefei 230026, China}
\affiliation{Hefei National Laboratory, University of Science and Technology of China, Hefei 230088, China}
\affiliation{Shanghai Research Center for Quantum Science and CAS Center for Excellence in Quantum Information and Quantum Physics, University of Science and Technology of China, Shanghai 201315, China}

\author{Xiao-Yu Dong}
\email{dongxyphys@ustc.edu.cn}
\affiliation{Hefei National Laboratory, University of Science and Technology of China, Hefei 230088, China}

\begin{abstract}
Intriguing phases may emerge when two-dimensional systems are coupled in a bilayer configuration. In particular, a Berezinskii-Kosterlitz-Thouless (BKT) paired superfluid phase was predicted and claimed to be numerically observed in a coupled $XY$ model with ferromagnetic interlayer interactions, as reported in [\href{https://doi.org/10.1103/PhysRevLett.123.100601}{Phys. Rev. Lett. 123, 100601 (2019)}]. However, both our Monte Carlo simulations and analytical analysis show that this model does not exhibit a BKT paired phase. We then propose a new model incorporating paired-phase gradient interlayer interactions to realize the BKT paired phase. Moreover, we observe that the anomalous magnetic dimension varies along the phase transition line between the disordered normal phase and the BKT paired phase. This finding requires an understanding beyond the conventional phase transition theory.
\end{abstract}

\maketitle

\section{introduction}  Coupling two layers of two-dimensional (2D) systems can give rise to exotic phases of matter that are absent in single-layer systems. These novel phases emerge from the interplay between interlayer coupling and the intrinsic properties of the individual layers, often leading to new collective behaviors and critical phenomena~\cite{cao2018unconventional, andrei2020graphene, doi:10.1126/science.aay5533, li2021quantum, PhysRevLett.128.026402, zhao2024realization, du2024atomic, meng2023atomic, wang2020localization, PARGA1980607, PhysRevLett.76.2910, PhysRevLett.110.146803, PhysRevLett.89.067001, BABAEV2004397, PhysRevB.88.220511, PhysRevA.90.043623, doi:10.1021/cr2003568, PhysRevLett.123.100601, PhysRevA.100.053604, PhysRevLett.128.195301, PhysRevB.111.094415, PhysRevX.14.011004, PhysRevLett.132.036502, PhysRevLett.128.157201, PhysRevB.106.184517, 
ma2025magnetic, PhysRevB.111.024511,grinenko2021state,shipulin2023calorimetric,halcrow2024probing}. A central question in this field is how the nature of interlayer coupling—whether linear, nonlinear, or multi-body—determines the hierarchy of emergent orders and their criticality. In this paper, we focus on the $XY$ model as a specific example to explore these effects.

In the single-layer case, the $XY$ model, which describes systems with U(1)-symmetric spins, undergoes the celebrated Berezinskii-Kosterlitz-Thouless (BKT) topological phase transition~\cite{berezinsky1972destruction, JMKosterlitz_1972, JMKosterlitz_1973, JMKosterlitz_1974, Kosterlitz_2016, Kardar2007}. This transition occurs between a low-temperature superfluid phase, characterized by the binding of vortex-antivortex pairs, and a high-temperature disordered phase, where these pairs unbind. The superfluid phase is characterized by algebraically decaying one-body correlations, reflecting quasi-long-range order (QLRO), while the disordered normal phase exhibits exponentially decaying correlations. The BKT transition plays a fundamental role in understanding critical phenomena across various physical systems, including ultracold atomic gases~\cite{hadzibabic2006berezinskii, Chomaz_2023} and optical lattices~\cite{PhysRevB.80.214513, schafer2020tools, zheng2025counterflow}. 

When two single-layer $XY$ models are coupled via interlayer interactions, new phases and transitions are anticipated. For example, recent work by Song and Zhang~\cite{PhysRevLett.128.195301} demonstrated that second-order Josephson coupling in a bilayer system induces an intermediate quasi-long-range ordered phase, corresponding to phase coherence of Cooper pairs (charge-4e superconductivity)~\cite{PhysRevLett.89.067001, BABAEV2004397,PhysRevB.88.220511, PhysRevB.107.064501, zeng2024high, PhysRevX.14.021025}. Victor Drouin-Touchette et al.'s work~\cite{drouin2022emergent} reports an emergent composite Potts order in the coupled hexatic-nematic $XY$ model. 

Our study is first motivated by the works \cite{PhysRevLett.123.100601, PhysRevB.111.094415}, where the two-body ferromagnetic interlayer interactions are introduced. They claimed that there is a novel BKT paired superfluid phase, sandwiched between the superfluid and disordered normal phases. In this BKT paired phase, the one-body correlations of spins within each layer decay exponentially, whereas a two-body correlation function of pairs of spins (one from the upper layer and one from the lower layer) exhibits a power-law decay, suggesting QLRO for paired spins. It were amazing if such a novel phase could emerge in this simple model where all interactions are ferromagnetic. Furthermore, this coupled layer model can be realized in ultracold atom systems, and a recent experimental work~\cite{rydow2024observation} has achieved a highly controllable bilayer of 2D Bose gases coupled via Josephson tunneling.

In this paper, we first reexamine the model presented in \cite{PhysRevLett.123.100601} using Monte Carlo simulations and demonstrate that, unfortunately, the BKT paired phase does not exist in this model. The experiment of 2D Bose gases~\cite{rydow2024observation} also supports our result. To realize a BKT paired phase, we propose a new model with paired-phase gradient interlayer couplings. This model exhibits three distinct phases: (1) a superfluid phase with two superfluids; (2) a disordered normal phase; and (3) a BKT paired phase that lies in between. Moreover, we observe that the anomalous magnetic dimension associated with the paired spin varies continuously along the phase boundary separating the BKT paired phase and the disordered normal phase.

\section{Main results} 
We consider extended $XY$ models on two coupled layers (labeled by $a$ and $b$) of 2D square lattices. The total Hamiltonian has the form 
\begin{eqnarray}
    H = H_a+H_b+H_{ab},
\end{eqnarray}
where $H_{\ell}= - \tilde{J} \sum_{\left\langle ij \right\rangle_{\ell}} \cos({\theta_{i,\ell} - \theta_{j,\ell}})$ with $\ell=a,b$ are the $XY$ intralayer interactions in $a$ and $b$ layer, respectively, with the same strength $\tilde{J}$. The variable $\theta_{i,\ell} \in (-\pi, \pi]$ represents the angle of the $XY$ spin in layer $\ell$ at site $i$, and $\left\langle ij \right\rangle_{\ell}$ denotes the nearest neighbors in layer $\ell$.

We consider two types of interlayer interactions $H_{ab}$. The first type is single-phase gradient interactions 
\begin{eqnarray}
    H_{ab}^{{\textrm{single}}} = - \widetilde{K} \sum_{i} \cos(\theta_{i,a} - \theta_{i,b}),
\end{eqnarray}
and the second type is paired-phase gradient interactions
\begin{eqnarray}
    H_{ab}^{ \textrm{pair}} = - \widetilde{K} \sum_{\left\langle ij \right\rangle} \cos(\theta_{i,a} + \theta_{i,b} - \theta_{j,a} - \theta_{j,b}),
\end{eqnarray}
where $\widetilde{K}\geq 0$. The corresponding total Hamiltonians are denoted as $H_{{\textrm{single}}}$ and $H_{\textrm{pair}}$, respectively. Considering their microscopic origin, $\theta$s are the phases of the underlying complex fields or bosonic operators. When amplitude fluctuations are suppressed, the low–energy Hamiltonian can be expressed solely in terms of these phases. The $H_{ab}^{\textrm{single}}$ is a gradient term between two single phases $\theta_{i,a}$ and $\theta_{i,b}$, and arises from $\phi^*_a(x)\phi_b(x)$ in field theory or, equivalently, $a_i^{\dag}b_i$ in bosonic lattice model. In contrast, $H_{ab}^{\textrm{pair}}$ is a gradient term between two paired phases $\theta_{i,a}+\theta_{i,b}$ and $\theta_{j,a}+\theta_{j,b}$, and originates from a gradient coupling $\phi^*_a(x)\phi_b^*(x)\phi_a(x')\phi_b(x')$ or paired hopping $a_i^{\dag}b_i^{\dag}a_jb_j$. The $H_{ab}^{\textrm{single}}$ can also be viewed as ferromagnetic interlayer interactions between the phases $\theta_{i,a}$ and $\theta_{i,b}$, since the energy is minimized when these two phases are aligned. In the following, we will use the dimensionless interaction coefficients $J = \widetilde{J}/k_BT$ and $K =\widetilde{K}/k_BT$ for convenience, where $k_B$ is the Boltzmann constant and $T$ is the temperature. In our analysis, two types of spin vectors $\mathbf{S}$ are considered. For a single-layer spin in layer $\ell$, the spin vector is defined as ${\bf S}^{\ell}_j= (\cos(\theta_{j,\ell}), \sin(\theta_{j,\ell}))$. In the coupled bilayer system, a paired spin vector is introduced as ${\bf S}^p_j = (\cos(\theta_{j,a} + \theta_{j,b}), \sin(\theta_{j,a} + \theta_{j,b}))$. For $H_{\textrm{single}}$, using the standard Swendsen-Wang (SW) cluster algorithm the critical slowing down is eliminated. For $H_{\textrm{pair}}$, we formulate a few variants of SW cluster methods, which help to greatly suppress the critical slowing down. Thus, extensive simulations can be performed for both systems.

\begin{figure}[!htbp]
    \centering
    \includegraphics[width=0.85\linewidth]{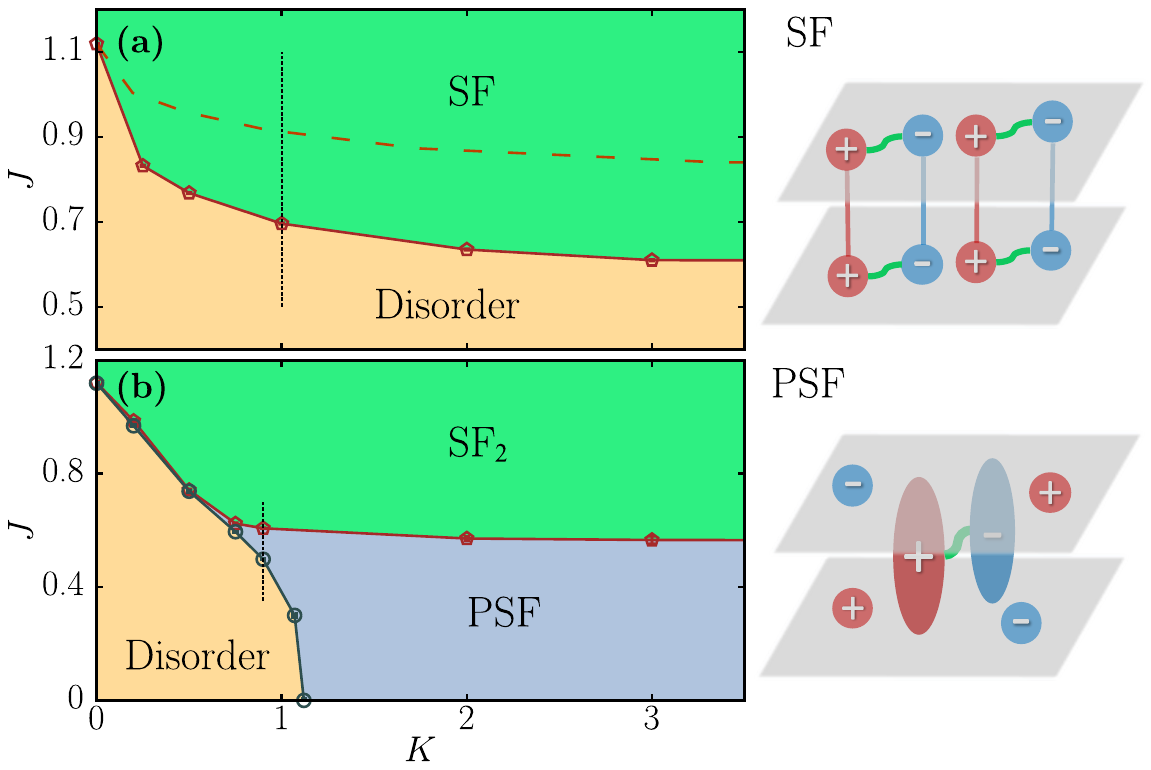}	
    \vspace*{-3mm}
    \caption{Phase diagram of (a) $H_{\textrm{single}}$ and (b) $H_{\textrm{pair}}$. The solid lines with data points on them are phase boundaries. The black dotted vertical lines correspond to the intervals used in Fig.~\ref{fig:model_1} and Fig.~\ref{fig:model_2}. ``SF" denotes the usual superfluid phase, {``SF$_2$" denotes the superfluid phase with two superfluids,}  ``Disorder" denotes the disordered normal phase, and ``PSF" denotes the BKT paired phase. The dashed orange line in (a) represents an additional phase transition claimed in \cite{PhysRevLett.123.100601, PhysRevB.111.094415}, which is not observed in our calculations. The schematic figures on the right half panel illustrate the key characteristics of the SF and PSF phases. In the SF phase, the vortices of the single-layer spins in each layer, as well as those of the paired spins, are tightly bound. The two vortices within a pair have the same sign due to the ferromagnetic interlayer interactions. In the PSF phase, the single-layer spins remain disordered, while the paired spins form bound vortices, giving rise to a superfluid of paired spins.}
    \label{fig:PD}
\end{figure}

The Hamiltonian $H_{\textrm{single}}$ is the same with that introduced in \cite{PhysRevLett.123.100601}, where it was argued that a novel BKT paired exists. We show both analytically and numerically that there is no such BKT paired phase, and the correct phase diagram has only two phases (a superfluid phase and a disordered normal phase) as presented in Fig.~\ref{fig:PD}(a). The phase boundary between the superfluid phase and the disordered phase is consistent with that obtained in \cite{PhysRevLett.123.100601}, while the other phase boundary reported in \cite{PhysRevLett.123.100601} (showed with dashed orange line) is absent. A possible reason why this incorrect phase boundary was obtained is analyzed in detail in the Appendix, see Fig.~\ref{fig:PD_M1_bug}. The phase diagram is determined with precision by the finite size scaling of $\xi_a$ and $\xi_p$, which are the second-moment correlation lengths corresponding to the spin vectors ${\bf S}^a$ and ${\bf S}^p$, respectively. Both $\xi_a$ and $\xi_p$ give the same phase transition points within numerical errors. Overall, the system $H_{\textrm{single}}$, of which the intra- and inter-layer interactions are both ferromagnetic, is essentially a 2D $XY$ model. This model has one and only one U(1) symmetry, i.e., the total energy remains unchanged if all spins in both layers are rotated by an arbitrary phase. According to the Ginzburg-Landau theory of phase transitions, it is natural that the $H_{\textrm{single}}$ has only one line of BKT transition. The inter-layer ferromagnetic interaction $K$ helps to reduce the critical coupling strength of $J$. In the $K \to \infty$ limit, the critical coupling $J_c$ becomes exactly half of $J_c (K=0)=1.119(2)$~\cite{PhysRevB.65.184405, komura2012a} for the single-layer case. 

The absence of the BKT paired phase can be further argued by comparing the spin-spin correlations of single-layer spin and paired spin. The one-body correlation function for the single-layer spin $\mathbf{S}^a$ is defined as $g_a(r) =  \langle {\bf S}^a_j\cdot {\bf S}^a_l\rangle =  \left \langle e^{i (\theta_{j,a} - \theta_{l,a})} \right \rangle$, and the two-body correlation function for the paired spin $\mathbf{S}^p$ is $g_p(r) = \langle {\bf S}^p_j\cdot {\bf S}^p_l\rangle=\left \langle e^{i (\theta_{j,a} + \theta_{j,b} - \theta_{l,a} - \theta_{l,b})} \right \rangle$, where $r$ is the distance between site $j$ and $l$ in the $xy$-plane. When $K = 0$, two layers are decoupled and the angles of spins in the two layers are independent, thus, we have $g_p(r)=g_a^2(r)$. In the limit $K\rightarrow \infty$, the ferromagnetic coupling between the two layers enforces the relative angle $\Delta_i = \theta_{i,a}-\theta_{i,b}$ to be zero. Using spin-wave theory~\cite{Kardar2007, nienhuis1984critical}, it can be shown that $g_p(r) = g_a^4(r)$. For finite $K$, the relative angle $\Delta_i\neq 0$ follows a Gaussian distribution, which introduces noise but preserves the scaling relation $g_p(r)\sim g_a^4(r)$ (verified numerically in the inset of  Fig.~\ref{fig:model_1}(c) at $K=1,J=0.8$, which is in the region of BKT paired phase reported in \cite{PhysRevLett.123.100601}). The anomalous magnetic dimension $\eta_a$ for single-layer spin and $\eta_p$ for paired spin are listed in Table~\ref{tab:model_1}. We can see that $\eta_p = 2\eta_a$ at $K=0$, while along the phase boundary with $K>0$ we have $\eta_p = 4\eta_a$, which are consistent with relation between the correlation functions. Furthermore, without resorting to any effective theory, we prove that in general $g_p(r)<g_a(r)$ in the limit $K\rightarrow \infty$ (see Appendix A for details). The BKT paired phase is characterized by exponential decaying $g_a(r)$ and algebraic decaying $g_p(r)$. However, our analysis shows that if $g_a(r)$ decays exponentially, $g_p(r)$ must decay even faster, ruling out the possibility of an algebraic decay for $g_p(r)$. The largest $K=3.5$ shown in the phase diagram is large enough to reflect the properties of $1/K\rightarrow 0$, since the value of $\cos(\theta_{i,a}-\theta_{i,b})$ increases rapidly with increasing $K$, which is $\sim 0.89$ at $K = 3.5$ (see Fig.~\ref{fig:corr_interlayer} in the Appendix). Increasing $1/K$ from $0$ to $1/3.5$, a new phase can emergence only if something extremely exotic happens, which is unlikely here since all interactions are trivially ferromagnetic.


To realize the BKT paired phase, we propose a new model $H_{\textrm{pair}}$ incorporating paired-phase gradient interlayer interactions. This term is fundamentally different from the effective ferromagnetic coupling in $H_{\textrm{single}}$, as it imposes no direct constraint on the relative angle between $\theta_{i,a}$ and $\theta_{i,b}$. 
The phase diagram of $H_{\textrm{pair}}$ is shown in Fig.~\ref{fig:PD}(b). Besides the superfluid and disordered phase, a BKT paired phase appears in between. The second-moment correlation lengths $\xi_a$ and $\xi_p$ give rise to two different phase transitions, which separate the BKT paired from the superfluid phase and the disordered phase, respectively. The BKT paired phase is characterized by exponential decaying $g_a(r)$ and algebraic decaying $g_p(r)$. The model $H_{\textrm{pair}}$ has a $\mathrm{U}(1)\times \mathrm{U}(1)$ symmetry, i.e., the total energy is unchanged if the spins in one of the layers are rotated by an arbitrary phase. Therefore, the Ginzburg-Landau theory admits two lines of phase transitions as observed in our simulations.

The existence of the BKT paired phase can be seen directly at the limit $K=+\infty$. Here, the $H_{\textrm{pair}}$ is dominated by the paired-phase gradient interactions, which is just the $XY$ model of paired spins ${\bf S}^p$, and $K=+\infty$ corresponds to the superfluid phase of paired spins. Fixing $J = 0$ and increasing $K$ from $K=0$ to $+\infty$, there must be a phase transition from the disordered phase to the BKT paired phase, and the critical couping is simply $K_c(J=0)=J_c(K=0)=1.119(2)$ for the single-layer $XY$ model. The terms with coefficients $J$ add interactions between single-layer spins in each layer. In the limit $J=+\infty$, the single-layer spins also form a superfluid in each layer. Therefore, there are three phases: (1) the disordered phase when both $K$ and $J$ are small, (2) the BKT paired phase, i.e., the superfluid of paired spin when $J$ is small and $K$ is large enough, (3) the superfluid phase with { two} superfluid components when $J$ is large enough. 

Another interesting point of the BKT paired phase is that along its phase boundary to the disordered phase, the anomalous magnetic dimension $\eta_p$ decreases continuously from $0.5$ to $0.25$ as $K$ increases (see Table~\ref{tab:model_2}). When $K=0$ and $J = 1.119(4)$, the two layers are decoupled, and the anomalous magnetic dimension is $\eta_p=0.5$. When $J=0$ and $K=1.12(1)$, $H_{\textrm{pair}}$ reduces to an $XY$ model of the paired spin, thus $\eta_p=0.25$, which is the same with the anomalous magnetic dimension of the BKT phase transition in a single-layer $XY$ model. The mechanism driving the continuous variation of the anomalous magnetic dimension along the phase boundary remains an open question.

\section{Algorithms and Observables}
For the Hamiltonian $H_{\textrm{single}}$, we employ the Swendsen-Wang (SW) algorithm~\cite{PhysRevLett.58.86, PhysRevLett.62.361} to update the configuration. The system size we simulate is up to $L=512$. To explore the Hamiltonian $H_{\textrm{pair}}$, we use a combination of various modified SW cluster algorithms and the Metropolis algorithm~\cite{Metropolis} to achieve high simulation efficiency and ensure the ergodicity of the configuration space (see Appendix C for details). The system size we simulate is up to $L=256$.

For a bilayer $XY$ spin system with $L \times L$ sites per layer and periodic boundary conditions, we sample the following observables. Each observable can be defined for both single-layer spins $\mathbf{S}^a$ and paired spins $\mathbf{S}^p$. In later discussions, subscripts will be used to distinguish between these two types of spins in the observables.

(a) The magnetization density, $M = L^{-2} \left|\sum_i \mathbf{S}_i\right|$. From this, the magnetic susceptibility is defined as $\chi = L^2 \langle M^2 \rangle$, where $\langle \cdot \rangle$ represents the statistical average.

(b) The Fourier transformation of the magnetization density, $M_k = L^{-2} \left|\sum_j \mathbf{S}_j e^{i \mathbf{k} \cdot \mathbf{r}_j}\right|$, where $\mathbf{r}_j$ is the coordinate of site $j$ and $\mathbf{k} = (2\pi / L, 0)$ is the smallest wave vector along the x-axis.

(c) The second-moment correlation length~\cite{PhysRevE.106.034138,PhysRevB.80.064418,PhysRevB.90.134420}, $\xi = \frac{1}{2\sin (|\mathbf{k}|/2)} \sqrt{\frac{\langle M^2 \rangle}{\langle M_k^2 \rangle} - 1}$. Moreover, the correlation-length ratio $\xi/L$ is an effective tool for identifying the critical points of phase transitions. In the disordered phase, where the correlation length $\xi$ is finite, this ratio drops to zero as the system size $L$ increases. In the QLRO phase, the ratio converges to a universal curve.

(d) The correlation function, $g(r) = \langle \mathbf{S}_0 \cdot \mathbf{S}_r \rangle = \langle e^{i(\theta_0 - \theta_r)} \rangle = \langle \cos(\theta_0 - \theta_r) \rangle$.

Additionally, we compute the magnetization and the correlation-function ratio, defined as $R_{M,n} = \langle M_a^2 \rangle^n / \langle M_p^2 \rangle$ and $R_{g,n} = g_a^n(r) / g_p(r)$ with integer $n$, respectively, to study the relation between the properties of single-layer spins and paired spins.


\begin{figure}[!htbp]
    \centering
    \includegraphics[width=0.85\linewidth]{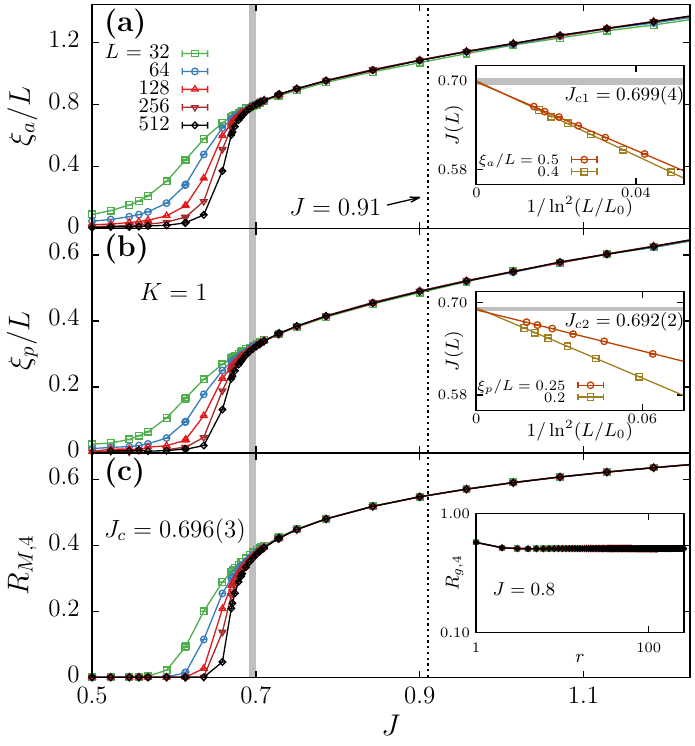}	
    \vspace*{-3mm}
    \caption{The numerical results for $H_{\textrm{single}}$ along the black dotted vertical line in Fig.~\ref{fig:PD}(a) with $K = 1$ are presented. The correlation length ratios $\xi_a/L$ for single-layer spins and $\xi_p/L$ for paired spins as functions of $J$ are shown in (a) and (b), respectively. In the corresponding inset, $J(L)$ is plotted against $1/\ln^2(L/L_0)$ for interpolation to estimate the critical point of the BKT transition, and the different colored lines represent different values of the correlation length ratio used for interpolation. The gray lines in the insets indicate the results from least squares fitting, which are consistent with the interpolation results. The gray line in the main figure indicates the transition point obtained by considering the ratios of two types of spins, while the dashed line represents another transition point reported in \cite{PhysRevLett.123.100601, PhysRevB.111.094415}. The magnetization ratio $R_{M,4}$ is plotted in (c), and its inset shows the correlation-function ratio $R_{g,4}(r)$ at $J=0.8$.}
    \label{fig:model_1}
\end{figure}

First, we show the numerical results for $H_{\textrm{single}}$. The Fig.~\ref{fig:model_1}(a) illustrates $\xi_a/L$ as a function of increasing $J$ along the black dotted line in the phase diagram shown in Fig.~\ref{fig:PD}(a), where $K=1$ is fixed. In the disordered phase at small $J$, the correlation length $\xi_a$ remains finite, leading to an inverse scaling of $\xi_a/L$ with system size $L$. In contrast, in the superfluid phase with quasi-long-range order, $\xi_a/L$ exhibits collapse across different values of $L$ due to finite size effects ($\xi_a$ diverges in the thermodynamic limit). The same analysis works for the $\xi_p/L$ in Fig.~\ref{fig:model_1}(b). The critical coupling $J_c$ at the phase transition point is determined by fitting the relation~\cite{PhysRevB.65.184405}
\begin{eqnarray}
    J(L) = J_c +\frac{\alpha}{(\ln L / L_0)^2} 
\end{eqnarray}
at a fixed $\xi/L$ in the disordered normal phase near the phase transition point, where $\alpha$ and $L_0$ are fitting parameters. The fitting of $\xi_a$ and $\xi_p$ are shown in the inset of Fig.~\ref{fig:model_1}(a) and (b), respectively. The corresponding critical coupling is found to be $J_{c_1}\approx 0.699(4)$ and $J_{c_2}\approx 0.692(2)$, whose values are the same with each other within numerical errors. The estimated critical strength is dramatically away from $J_c(K=1)=0.91$ for the phase boundary between the superfluid phase and the BKT paired phase in \cite{PhysRevLett.123.100601}. As shown in Fig.~\ref{fig:model_1}(a)(b), for $J \approx 0.696$, the correlation-length ratios, $\xi_{a(p)}/L$, quickly conserve to a smooth function for large systems, and do not display any singular behavior around $J =0.91$ that was marked by the arrow in Fig.~\ref{fig:model_1}(a). This indicates that at the phase transition from the disordered normal phase to the superfluid phase, quasi-long-range order emerges simultaneously for both the single-layer spins and the paired spins. These results provide strong and unambiguous evidence that the BKT paired phase reported in \cite{PhysRevLett.123.100601} does not exist.  

To further support this conclusion, we directly compare the squared magnetization density and correlation functions of $\mathbf{S}^a$ and $\mathbf{S}^p$. The ratio of two types of squared magnetization densities, $R_{M,4}$, is shown in Fig.~\ref{fig:model_1}(c). In the disordered normal phase, both $\langle M_a^2 \rangle$ and $\langle M_p^2 \rangle$ decrease to zero exponentially as $L$ increases. In the superfluid phase with QLRO, the relation $\langle M_a^2 \rangle^4\sim \langle M_p^2 \rangle$ holds. This behavior is consistent with that of the spatial correlations of $\mathbf{S}^a$ and $\mathbf{S}^p$. In the superfluid phase, the correlation functions have the relation $g_a^4(r)\sim  g_p(r) $, as shown in the inset of Fig.~\ref{fig:model_1}(c) where $R_{g,4}(r)$ is plotted for a representative point in the superfluid phase ($K = 1$, $J = 0.8$). 

\begin{table}[!ht]
    \centering
    \caption{For the $H_{\textrm{single}}$, the values of  $J_{c1}({\rm single})$ from the single-layer correlation-length ratio agree well with $J_{c2}({\rm paired})$ from the paired correlation-length ratio. For the decoupled case ($K=0$), the paired and the single-layer exponents are related as $\eta_p = 2 \eta_a=1/2$, while for $K >0$, the relation reads $\eta_p = 4 \eta_a=1$. These are well supported by the numerical results.}
    \vspace{20pt}
    \begin{tabular}{lll|ll}
    \hline
    \hline
        $K_c$ & $J_{c1}({\rm single})$ & $J_{c2}({\rm paired})$ & $\eta_a$ & $\eta_p$ \\ \hline
        0    & 1.121(5) & 1.119(4) & 0.252(9)  & 0.51(1)  \\ 
        0.25 & 0.840(4) & 0.830(2) & 0.252(2)  & 1.002(1) \\ 
        0.50 & 0.774(4) & 0.766(2) & 0.2516(9) & 0.999(2) \\ 
        1.00 & 0.699(4) & 0.692(2) & 0.2520(4) & 0.998(1) \\ 
        2.00 & 0.636(3) & 0.632(2) & 0.2507(5) & 0.997(2) \\ 
        3.00 & 0.606(3) & 0.604(3) & 0.2519(6) & 0.999(3) \\ \hline\hline
    \end{tabular}
    \label{tab:model_1}
\end{table}

Representative points on the phase boundary are summarized in Table~\ref{tab:model_1}. The corresponding anomalous magnetic dimensions are obtained by fitting the relation~\cite{PhysRevB.37.5986, PhysRevB.55.3580, KENNA1997583}
\begin{align}
    \chi = L^{2-\eta}(\ln L + C_1)^{-2\hat{\eta}}(a_0 + b_1 L^{-\omega}) 
    \label{eq:fitting_eta_ln}
\end{align} 
near the phase transition points, where $\chi$ denotes the magnetic susceptibility
and $ a_0, b_1, C_1 $ are fitting parameters, and $ L^{-\omega} $ represents the finite-size correction term. Here, $\hat{\eta} = -\eta/4$ is fixed due to the renormalization analysis of the BKT phase transition~\cite{Kosterlitz_2016}. The values of the other fitting parameters are provided in the Appendix D. Along the phase boundary for $K>0$, the exponents $\eta_a$ and $\eta_p$, corresponding to $\mathbf{S}^a$ and $\mathbf{S}^p$, are approximately fixed at $\eta_a\approx 0.25$ and $\eta_p \approx 1.0$, respectively. The value of $\eta_a$ aligns with well-established results for the single-layer BKT phase transition. Since $\langle M_a^2 \rangle^4\sim \langle M_p^2 \rangle$, we find that in the superfluid phase $\eta_p \approx 4\eta_a$. The limit $K = 0$ is special, as the two layers decouple in this case. At this point, we have $g_p(r)=g_a^2(r)$, which leads to $\eta_p = 2\eta_a$.

\begin{figure}[!htbp]
    \centering
    \includegraphics[width=0.83\linewidth]{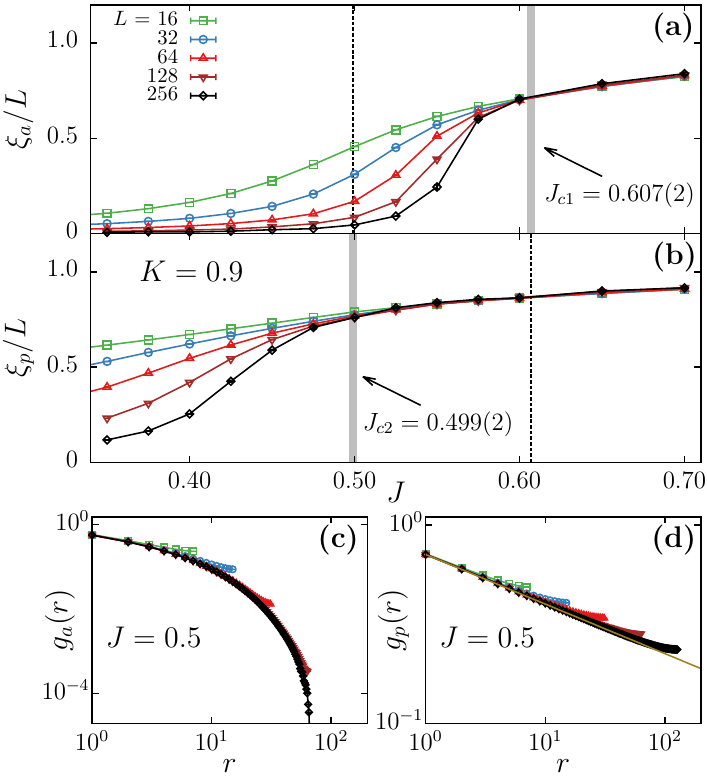}	
    \vspace*{-3mm}
    \caption{The numerical results for $H_{\textrm{pair}}$ along the black dotted vertical line in Fig.~\ref{fig:PD}(b) with $K = 0.9$ are presented. The correlation length ratios $\xi_a/L$ for single-layer spins and $\xi_p/L$ for paired spins as functions of $J$ are shown in (a) and (b), respectively. The gray lines in (a) and (b) represent the estimated transition points for single-layer spins and paired spins, respectively.
    The correlation functions $g_a(r)$ and $g_p(r)$ at $J = 0.5$, effectively at the paired BKT point $J_{c1}=0.499(2)$, are plotted in (c) and (d). It is clearly shown that the two-point correlation function $g_a(r)$ within a single layer decays exponentially fast, while the paired correlation $g_p(r)$ decays algebraically. Note that $g_a$ is significantly smaller than $10^{-4}$ for $r \approx 50$ while $g_p \approx 0.2$ for $r = 128$.}
    \label{fig:model_2}
\end{figure}

Then, we show the numerical results for $H_{\textrm{pair}}$. The values of second-momentum correlation length $\xi_a$ and $\xi_p$ along the black dotted line with fixed $K=0.9$ in Fig.~\ref{fig:PD}(b) are plotted in Fig.~\ref{fig:model_2}(a) and (b), respectively. At small $J$ in the disordered phase, both $\xi_a$ and $\xi_p$ are finite. As $J$ increases past $J_{c2}\approx 0.499(2)$, $\xi_p/L$ collapses across different system sizes $L$, indicating the onset of QLRO of $\mathbf{S}^p$. This is consistent with the power-law decay of $g_p(r)$ shown Fig.~\ref{fig:model_2}(c) at $J=0.5$. In contrast, $\xi_a$ remains finite and $g_a(r)$ decays exponentially (Fig.~\ref{fig:model_2}(d)) until $J$ reaches $J_{c1}\approx 0.607(2)$. The intermediate region between $J_{c1}$ and $J_{c2}$ corresponds to the BKT paired phase. Beyond $J_{c1}$, the system enters the superfluid phase, where both $\xi_a/L$ and $\xi_p/L$ exhibit collapse for different $L$, and both correlation functions decay with power-law behavior. 

The superfluid phase in $H_{\textrm{pair}}$ also exhibits a distinct behavior compared to that in $H_{\textrm{single}}$. In $H_{\textrm{single}}$, the effective ferromagnetic coupling between the two layers results in a finite and rapidly increasing value of $\langle \cos(\theta_{i,a}-\theta_{i,b})\rangle$ as $K$ increases. This coupling strongly aligns $\mathbf{S}^a$ and $\mathbf{S}^b$, indicating that the superfluid states in two layers are not independent. This alignment is precisely why the QLRO of the single-layer and paired spins emerges simultaneously. In contrast, the paired-phase gradient interactions in $H_{\textrm{pair}}$ do not impose any preference on the relative angle between $\mathbf{S}^a$ and $\mathbf{S}^b$, leading to $\langle \cos(\theta_{i,a}-\theta_{i,b})\rangle=0$ (see Fig.~\ref{fig:corr_interlayer_paired} in the Appendix B).  { Indeed, from the perspective of symmetry, the $\mathrm{U}(1)\times \mathrm{U}(1)$ symmetry corresponds to two independent superfluid modes: the phase sum $\theta_+ = \theta_a + \theta_b$ and the phase difference $\theta_- = \theta_a - \theta_b$. In the disordered phase, both modes are disordered. In the paired BKT phase, $\theta_+$ exhibits QLRO while $\theta_-$ remains disordered. At even lower temperatures, both $\theta_+$ and $\theta_-$ develop QLRO. This is why we label this phase with the subscript $\mathrm{SF}_2$ in Fig.~\ref{fig:PD}(b).}

\begin{figure}[!htbp]
    \centering
    \includegraphics[width=0.85\linewidth]{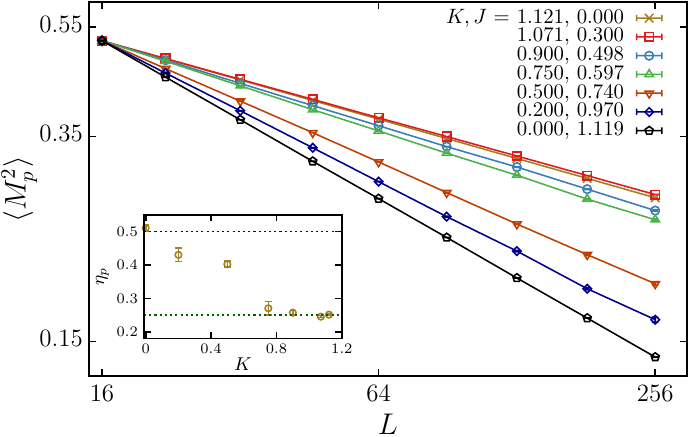}	
    \vspace*{-3mm}
    \caption{The log-log plot of the squared magnetization density $ \langle M_p^2 \rangle$ for paired spins versus system size $ L $ at various critical points along the phase boundary between the BKT paired phase and the disordered phase. The approximately straight lines with different slopes clearly indicate that the paired anomalous magnetic exponent $\eta_p$ varies along the phase boundary. This is in contrast with the naive expectation from the universality that it should be a constant,  raising an open question on the underlying mechanism. The inset displays $\eta_p$ versus increasing $K$ along the phase boundary.}
    \label{fig:eta_p_model_2}
\end{figure}

\begin{table}[!ht]
    \centering
    \caption{Estimates of critical points $J_{c2}$ and exponents $\eta_p$ for paired spins on the phase boundary between disordered normal phase and BKT paired phase for $H_{\textrm{pair}}$.}
    \vspace{20pt}
    \begin{tabular}{ll|l}
    \hline
    \hline
        $K_c$   & $J_{c2}({\rm paired})$  & $\eta_p$ \\ \hline
        0       & 1.119(4)    & 0.51(1)  \\ 
        0.20    & 0.969(4)    & 0.43(2)  \\ 
        0.50    & 0.738(4)    & 0.402(9) \\ 
        0.75    & 0.595(2)    & 0.27(2)  \\ 
        0.90    & 0.499(2)    & 0.257(7) \\ 
     1.07(1)    & 0.30        & 0.245(3) \\
     1.12(1)    & 0           & 0.251(2) \\ \hline\hline
    \end{tabular}
    \label{tab:model_2}
\end{table}

Along the phase boundary between the BKT paired phase and the disordered phase, Fig.~\ref{fig:eta_p_model_2} shows the changes of slope of the log-log plot of $\langle M_p^2\rangle$ versus $L$, since $\langle M_p^2\rangle\sim L^{-\eta_p}$. Detailed numerical values of $\eta_p$ are provided in Table~\ref{tab:model_2} and visualized in the inset of Fig.~\ref{fig:eta_p_model_2}.

\section{Conclusion and Discussions}  We investigate the emergence of a BKT paired superfluid phase in two bilayer $XY$ models, $H_{\textrm{single}}$ and $H_{\textrm{pair}}$, using extensive Monte Carlo simulations. Our results reveal that the BKT paired phase is absent in $H_{\textrm{single}}$, contrary to findings in previous studies~\cite{PhysRevLett.123.100601, PhysRevB.111.094415}. In this model, the interlayer ferromagnetic interactions lead to the simultaneous establishment of QLRO for both single-layer spins and paired spins, which is the physical reason underlying the absence of BKT paired phase. We propose a new model $H_{\textrm{pair}}$ with paired-phase gradient interlayer interactions, and demonstrate the existence of the BKT paired phase in this model. The paired-phase gradient interactions do not constrain the relative angles between spins in the upper and lower layers. Thus, the QLRO can be set up only in the paired spins in a certain region of the phase diagram. We also observe that the phase transition between the disordered normal phase and the BKT paired phase is quite unusual, as the $\eta_p$ varies continuously along the phase boundary. This behavior lies beyond the conventional understanding of critical lines. The continuous variation of the $\eta_p$ may be an intrinsic feature of the model, potentially explained by renormalization effects in the underlying spin-wave theory. Further analytical and numerical studies will be necessary to fully elucidate the nature of this critical behavior. Finally, we mention that by generalizing the Hamiltonian, $H_{\textrm{pair}}$, to higher dimensions, a phase diagram similar to Fig.~\ref{fig:PD}(b) should be observed, and the BKT paired phase becomes the paired superfluidity of long-range order.

\section*{Acknowledgments}
This work has been supported by the National Natural Science Foundation of China (Grant No. 12275263), the Innovation Program for Quantum Science and Technology (Grant No. 2021ZD0301900), and the Natural Science Foundation of Fujian Province of China (Grant No. 2023J02032).

{\section*{data availability}
The data that support the findings of this article are openly available~\cite{tensofermi2025data}.
}

\onecolumngrid
\appendix


\section{Relation between $g_a(r)$ and $g_p(r)$ in the $H_{\textrm{single}}$ model}
\label{sec:derive}
In this Appendix, we derive the relation between the correlation function $g_a(r)$ of single-layer spins and the correlation function $g_p(r)$ for paired spins in $H_{\textrm{single}}$ model. We focus on the case with coupling strength $K \geq 0$, and provide corresponding numerical results that are consistent with the derivation. Additionally, we prove that $g_p(r) \leq g_a(r)$ in the $K \to \infty$ limit, which is crucial for the argument that the paired phase is absent in the ferromagnetic coupling model. Finally, we discuss the incorrect phase boundary in Refs.~\cite{PhysRevLett.123.100601, PhysRevB.111.094415}.

\subsection{The case at $K=0$}
When $K=0$, two layers are decoupled, and the angles between spins in two layers are independent with each other. Hence, the paired correlation function $g_p(r)$ is the square of the single-layer correlation function $g_a(r)$:
\begin{align}
        g_p(r) = & \left \langle e^{i (\theta_{0,a} + \theta_{0,b} - \theta_{r,a} - \theta_{r,b})} \right \rangle 
        \nonumber \\ = & \left \langle e^{i (\theta_{0,a} - \theta_{r,a})} e^{i(\theta_{0,b}- \theta_{r,b})} \right \rangle 
        \nonumber \\ = & \left \langle e^{i (\theta_{0,a} - \theta_{r,a})} \right \rangle \left\langle e^{i(\theta_{0,b}- \theta_{r,b})} \right \rangle 
        \nonumber \\ = & ~ g_a^2(r).
\end{align}

Therefore, if the system is in a phase with QLRO and we denote the single-layer anomalous magnetic dimension as $\eta_a = \eta$, we can derive, based on the characteristic power-law decay of correlations in this phase~\cite{Kardar2007}, that
\begin{align}
    g_p(r) = g_a^2(r) \sim (r^{-\eta})^2 = r^{-\eta_p}.
\end{align}
Thus, the paired anomalous magnetic dimension is $\eta_p = 2 \eta$. Note that, for simplicity, we ignore the logarithmic correction exponent $\hat{\eta}$ here.

\begin{figure}[!htbp]
    \centering
    \includegraphics[width=0.6\linewidth]{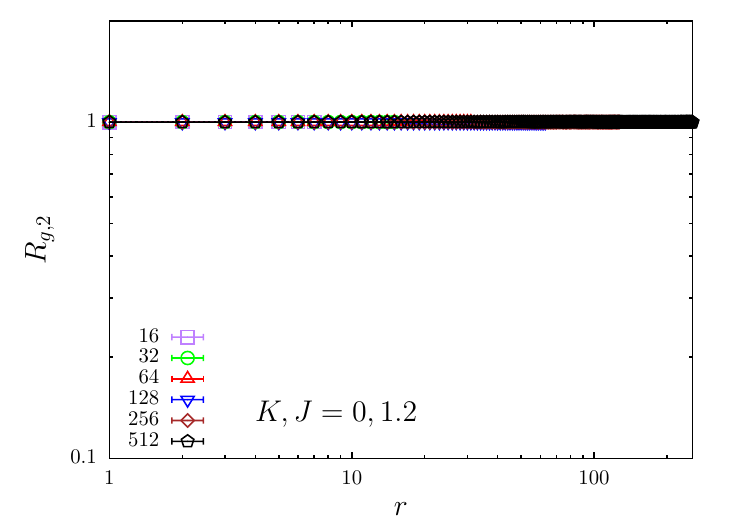}	
    \vspace*{-3mm}
    \caption{The ratio of two types of correlation functions $R_{g,2}$ at $K=0$ and $J=1.2$ (QLRO phase) for the $H_{\textrm{single}}$ model. The straight-line behavior and collapse at $R_{g,2} = 1$ indicate that $g_p(r) = g_a^2(r)$. }
    \label{fig:K_0_gr}
\end{figure}

As shown in Fig.~\ref{fig:K_0_gr}, in the QLRO phase, the ratio of the two types of correlation functions $R_{g,2}(r) = g_a^2(r)/g_p(r)$ exhibits a straight line and shows good collapse at $R_{g,2} = 1$. This clearly indicates that $g_p(r) = g_a^2(r)$.

\subsection{The case at $K\rightarrow\infty$ limit}
Before considering the case for $K > 0$, let us first examine the $K \to \infty$ limit for simplicity. In this limit, the strong ferromagnetic couplings between two layers force the angles of spin to align, i.e., $\theta_a = \theta_b$. Thus, we denote $\theta_{0,a} = \theta_{0,b} = \theta_{0}$ and $\theta_{r,a} = \theta_{r,b} = \theta_{r}$. Therefore, the correlation functions of single-layer spins and paired spins can then be written as
\begin{align}
g_a(r) =& \left\langle e^{i (\theta_{0} - \theta_{r})} \right\rangle, \\
g_p(r) =& \left\langle e^{i (2\theta_{0} - 2\theta_{r})} \right\rangle.
\end{align}

According to spin-wave theory~\cite{Kardar2007,nienhuis1984critical}, at low temperatures, the cost of small fluctuations around the ground state is obtained by a quadratic expansion, which gives $\frac{J}{2} \int d^2 {\bf x} (\nabla \theta)^2$ in the continuum limit, where $J$ is the coupling strength. Therefore, in two dimensions, the standard rules of Gaussian integration yield
\begin{align}
\left\langle e^{i (\theta_{0} - \theta_{r})} \right\rangle = e^{-\frac{1}{2} \left\langle (\theta_{0} - \theta_{r})^2 \right\rangle} = e^{-\frac{1}{2\pi J} \ln \left(\frac{r}{a}\right)} = \left(\frac{r}{a}\right)^{-\frac{1}{2\pi J}},
\end{align}
where $a$ is a short-distance cutoff. For a lattice, we set $a = 1$. Hence, the anomalous magnetic dimension $\eta$ can be extracted as $\frac{1}{4}$ when considering the BKT critical point $J_c = \frac{2}{\pi}$, obtained from renormalization group theory~\cite{Kardar2007}.

For the paired correlation function $g_p(r)$, we immediately find that
\begin{align}
\left\langle e^{i (2\theta_{0} - 2\theta_{r})} \right\rangle = e^{-\frac{1}{2} \left\langle (2\theta_{0} - 2\theta_{r})^2 \right\rangle} = e^{- 2\left\langle (\theta_{0} - \theta_{r})^2 \right\rangle} = \left(\frac{r}{a}\right)^{-\frac{2}{\pi J}}.
\end{align}
Thus, we obtain $g_p(r) = g_a^4(r)$ and $\eta_p = 4\eta$ for $K\to\infty$. 

\subsection{The case at finite positive $K$}
For the case $K > 0$, at low temperatures, we can apply the spin-wave approximation to the ferromagnetic coupling between two layers. This leads to the relation $\theta_{j,a} - \theta_{j,b} = \Delta_j$, where $\Delta_j$ follows a Gaussian distribution, i.e., $\Delta_j \sim N(0, \sigma)$.
This notation means that $\Delta_j$ is normally distributed with a mean 0 and a standard deviation $\sigma$.

Based on this, we obtain the following relations: 
\begin{align} \theta_{0,a} + \theta_{0,b} = \Delta_0 + 2 \theta_{0,b} = \Delta_0 + 2 \theta_{0} \quad \text{and} \quad \theta_{r,a} + \theta_{r,b} = \Delta_r + 2 \theta_{r,b} = \Delta_r + 2 \theta_{r}. \end{align}

Hence, the paired correlation function can be written as 
\begin{align} g_p(r) = \left\langle e^{i (2\theta_{0} + \Delta_0 - 2\theta_{r} - \Delta_r)} \right\rangle = \left\langle e^{i (2\tilde{\theta}_{0} - 2\tilde{\theta}_{r})} \right\rangle, 
\end{align} 
where $\tilde{\theta}_j = \theta_j + \Delta_j / 2$.

When $K \to \infty$, the variance of the Gaussian distribution $\sigma \to 0$, and therefore the distribution of $\Delta_j$ tends to a delta function $\delta(0)$, which implies $\Delta_j = 0$. In this case, we have $\theta_{j,a} = \theta_{j,b}$, as discussed in the previous subsection.

However, when $K$ is finite, this can be interpreted as applying Gaussian noise $\Delta_j / 2$ to the angles. This noise just affects the amplitude but does not affect the scaling behavior as $g_p(r) \sim g_a^4(r)$, so the relation $\eta_p = 4 \eta$ still holds.

 \begin{figure}[!htbp]
    \centering
    \includegraphics[width=0.6\linewidth]{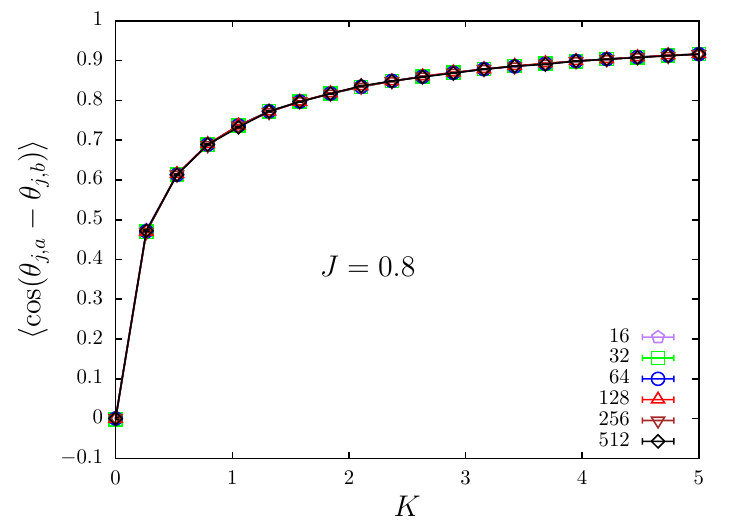}	
    \vspace*{-3mm}
    \caption{The inter-layer correlation $\langle \cos (\theta_{j,a} - \theta_{j,b})  \rangle$ versus coupling strength $K$ at $J = 0.8$ for the ferromagnetic coupling model. } 
    \label{fig:corr_interlayer}
\end{figure}

 \begin{figure}[!htbp]
    \centering
    \includegraphics[width=0.6\linewidth]{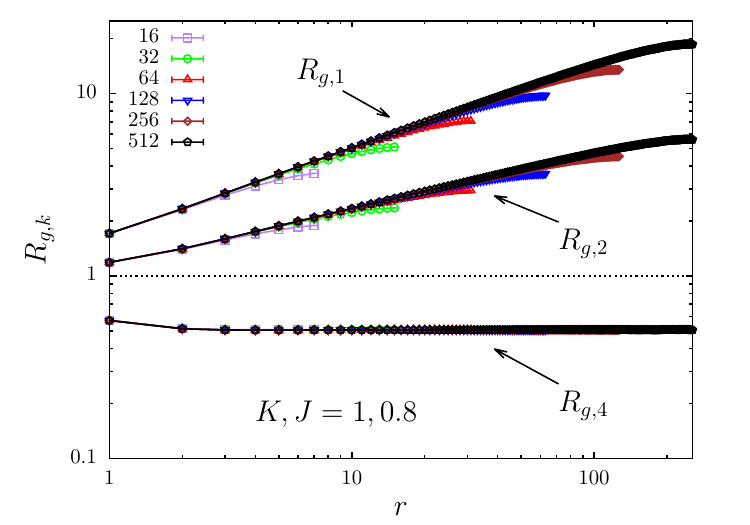}	
    \vspace*{-3mm}
    \caption{The ratio of two types of correlation functions $R_{g,k}$ for $k=1,2,4$ at $K=1$ and $J=0.8$ (QLRO phase) for the ferromagnetic coupling model.}
    \label{fig:inequality_showq}
\end{figure}

Numerically, as shown in Fig.~\ref{fig:corr_interlayer}, 
we measure the inter-layer correlation $\left\langle e^{i\Delta_j}\right\rangle = \left\langle \cos(\theta_{j,a} - \theta_{j,b}) \right\rangle$. It rapidly increases to nearly 0.9 as $K$ increases beyond $3$. This indicates that the properties of the system rapidly approach the case of $K \to \infty$.
Moreover, in Fig.~\ref{fig:inequality_showq}, the good data collapse of the ratio of the correlation functions $R_{g,4}$ clearly indicates that the relation $g_p(r) \sim g_a^4(r)$ holds for finite $K$.

\subsection{The absence of BKT paired phase}

Here, we aim to prove that $g_p(r) \leq g_a(r)$ in the $K \to \infty$ limit, i.e., $\theta_{a} = \theta_{b}$. Note that if this inequality holds and $g_a(r)$ decays exponentially, then $g_p(r)$ must decay even faster. Therefore, it is impossible for $g_p(r)$ to exhibit an algebraic decay behavior, implying the absence of the so-called paired BKT phase proposed in \cite{PhysRevLett.123.100601}.

Mathematically, it is easy to check that this proposition is equivalent to the following inequality:
\begin{align}
    \int_{-\pi}^{\pi} f(x) \cos(2x) \, \mathrm{d}x \leq \int_{-\pi}^{\pi} f(x) \cos(x) \, \mathrm{d}x,
\end{align}
where $x = \theta_0 - \theta_r \in (-\pi, \pi]$ and the function $f(x)$ is the distribution function of $x$.  

Considering the U(1) symmetry of the spins and the ferromagnetic interactions within each layer, the distribution function $f(x)$ is normalized ($\int_{-\pi}^{\pi} f(x) \, \mathrm{d}x = 1$), non-negative ($f(x) \geq 0$ for all $x \in (-\pi, \pi]$), even ($f(x) = f(-x)$), and monotonically decreasing in $[0, \pi]$.

Since both $f(x)$ and $\cos(nx)$ ($n=1,2$) are even functions, the integral over $(-\pi, \pi]$ can be expressed as the integral over $[0, \pi]$:
\begin{align}
\int_{0}^{\pi} f(x) \cos(2x) \, \mathrm{d}x \leq \int_{0}^{\pi} f(x) \cos(x) \, \mathrm{d}x.
\end{align}

Next, define the difference function $h(x) = \cos(x) - \cos(2x)$. Our goal is to show that:
\begin{align}
\int_{0}^{\pi} f(x) h(x) \, \mathrm{d}x \geq 0.
\end{align}

It is straightforward to observe that $h(x) \geq 0$ for $x \in [0, \frac{2\pi}{3}]$ and $h(x) \leq 0$ for $x \in \left(\frac{2\pi}{3}, \pi\right]$. Additionally, the following property holds:
\begin{align}
    \int_{0}^{\frac{2\pi}{3}} h(x) \, \mathrm{d}x = - \int_{\frac{2\pi}{3}}^{\pi} h(x) \, \mathrm{d}x = \frac{3\sqrt{3}}{4}.
\end{align}

Thus, we can split the integral into two parts, yielding:
\begin{align}
\int_{0}^{\frac{2\pi}{3}} f(x) h(x) \, \mathrm{d}x \geq - \int_{\frac{2\pi}{3}}^{\pi} f(x) h(x) \, \mathrm{d}x.
\end{align}

To prove the inequality, we show that the minimum of the left-hand side is greater than or equal to the maximum of the right-hand side. Since $f(x)$ is monotonically decreasing in $[0, \pi]$, we have the following:

For the left integral, on $[0, \frac{2\pi}{3}]$, we have $f(x) \geq f\left(\frac{2\pi}{3}\right)$:
\begin{align}
\int_{0}^{\frac{2\pi}{3}} f(x) h(x) \, \mathrm{d}x \geq f\left(\frac{2\pi}{3}\right) \int_{0}^{\frac{2\pi}{3}} h(x) \, \mathrm{d}x = f\left(\frac{2\pi}{3}\right) \cdot \frac{3\sqrt{3}}{4}.
\end{align}

For the right integral, on $[\frac{2\pi}{3}, \pi]$, we have $f(x) \leq f\left(\frac{2\pi}{3}\right)$:
\begin{align}
-\int_{\frac{2\pi}{3}}^{\pi} f(x) h(x) \, \mathrm{d}x \leq -f\left(\frac{2\pi}{3}\right) \int_{\frac{2\pi}{3}}^{\pi} h(x) \, \mathrm{d}x = f\left(\frac{2\pi}{3}\right) \cdot \frac{3\sqrt{3}}{4}.
\end{align}

Therefore, the inequality $g_p(r) \leq g_a(r)$ is proven. Moreover, as shown in Fig.~\ref{fig:inequality_showq}, the ratio $R_{g,1} > 1$ indicates that $g_p(r) < g_a(r)$, which is consistent with our derivation. Furthermore, the ratio $R_{g,2} > 1$ suggests that $g_p(r) < g_a^2(r)$, i.e., $g_p(r)$ decays faster than anticipated here.

\subsection{Analysis of the incorrect phase boundary in Refs.~\cite{PhysRevLett.123.100601, PhysRevB.111.094415}}

A key property of the $H_{\textrm{single}}$ model is that, in the low-temperature superfluid phase, the relation $\eta_p = 2 \eta$ holds for $K = 0$, and $\eta_p = 4 \eta$ holds for $K > 0$. These relations are supported by both theoretical derivation and numerical results presented above and in the main text.

\begin{figure}[!htbp]
    \centering
    \includegraphics[width=0.6\linewidth]{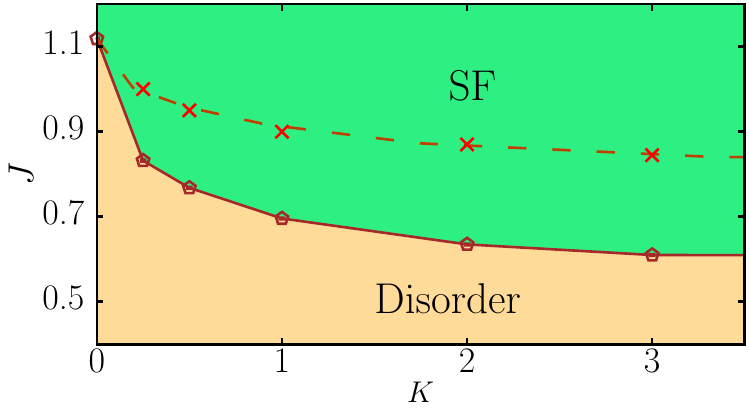}	
    \vspace*{-3mm}
    \caption{Phase diagram of the $H_{\textrm{single}}$ model. The red crosses indicate points where the paired anomalous magnetic dimension satisfies $\eta_p = \frac{1}{2}$, which is close to the phase boundary reported in Refs.~\cite{PhysRevLett.123.100601, PhysRevB.111.094415}.} 
    \label{fig:PD_M1_bug}
\end{figure}

At the BKT phase transition point, it is known that $\eta = \frac{1}{4}$, which leads to $\eta_p = \frac{1}{2}$ for $K = 0$ and $\eta_p = 1$ for $K > 0$. As the temperature decreases (or equivalently, as $J$ increases), the exponent $\eta$ monotonically decreases to zero.
Therefore, as a conjecture, if one incorrectly assumes that $\eta_p = \frac{1}{2}$ at the critical point not only for $K = 0$ but also for $K > 0$, this would result in a phase boundary that lies above the true BKT transition boundary (solid line in Fig.~\ref{fig:PD_M1_bug}).

Hence, based on the scaling relation $\langle M_p^2 \rangle \sim L^{-\eta_p}$, we extract the values of $J$ that satisfy $\eta_p = \frac{1}{2}$ for different $K = 0.25, 0.5, 1, 2, 3$, and the corresponding points are marked with red crosses in Fig.~\ref{fig:PD_M1_bug}. Interestingly, these points lie close to the phase boundary reported in Refs.~\cite{PhysRevLett.123.100601, PhysRevB.111.094415} (see dashed line in Fig.~\ref{fig:PD_M1_bug}). 

We therefore attribute the previously reported, but incorrect, phase boundary to a failure to distinguish between the distinct scaling behaviors of $\eta_p$ for $K = 0$ and $K > 0$.

\section{Inter-layer correlation for the $H_{\textrm{pair}}$ model}
\begin{figure}[!htbp]
    \centering
    \includegraphics[width=0.6\linewidth]{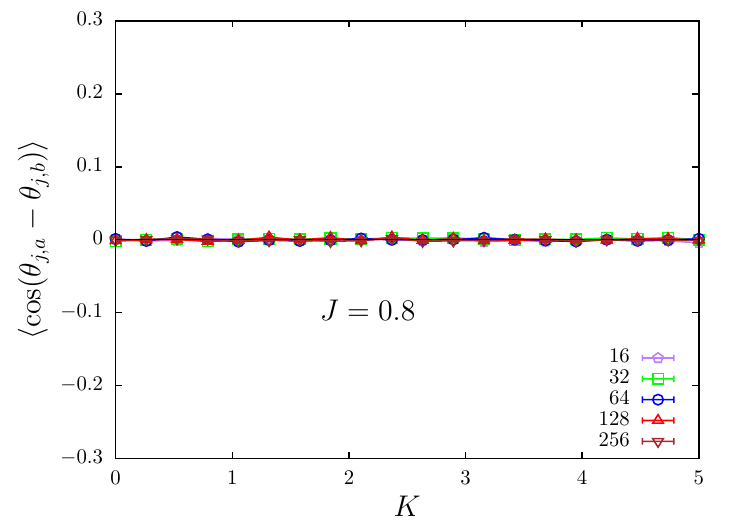}	
    \vspace*{-3mm}
    \caption{The inter-layer correlation $\langle \cos (\theta_{j,a} - \theta_{j,b})  \rangle$ versus coupling strength $K$ at $J = 0.8$ for the $H_{\textrm{pair}}$ model. } 
    \label{fig:corr_interlayer_paired}
\end{figure}

{
In Fig.~\ref{fig:corr_interlayer_paired}, we observe that, in contrast to the ferromagnetic coupling model shown in Fig.~\ref{fig:corr_interlayer}, the {paired-phase gradient coupling} model does not exhibit ferromagnetic interlayer correlations as $K$ increases; that is, $\langle \cos(\theta_{j,a} - \theta_{j,b}) \rangle = 0$. This behavior can be understood by considering the constraint that emerges in the limit $K \to \infty$, where $\theta_{i,a} + \theta_{i,b} = \theta_{j,a} + \theta_{j,b} = \phi$, implying that the sum of phases on each site couples to a global variable $\phi$ via the paired-phase gradient interaction.

Consequently, the individual phases in the two layers satisfy $\theta_{i,a} = \phi - \theta_{i,b}$, so that for any given configuration, the two layers are related through a fixed phase difference. However, since $\phi$ varies across configurations, the averaged interlayer phase correlation vanishes.

}

\section{Cluster algorithm for bilayer models}
For a standard cluster algorithm~\cite{PhysRevLett.58.86, PhysRevLett.62.361}, there are two main steps: cluster formation and spin operation. In the cluster formation step, clusters are formed by placing bonds between interacting lattice sites with probability $\max[0, P_b]$. In the spin operation step, operations are performed on the spins within these formed clusters.

Specifically, consider a two-body interaction between site $i$ and site $j$, with the energy unit denoted as $\varepsilon_{ij}$. The partition function of this system can be written (where the inverse temperature $\beta$
is absorbed into $\varepsilon_{ij}$) as 
\begin{align} 
\mathcal{Z} = \sum_{s} \prod_{ij} e^{-\varepsilon_{ij}}, 
\end{align} 
where $s$ represents all possible configurations. For a given configuration, the product of all interaction unit weights is $\prod_{ij} e^{-\varepsilon_{ij}}$. Notably, if the energy level of the unit $\varepsilon_{ij}$ is binary—taking values $\varepsilon_0$ for the lower energy level and $\varepsilon_1$ for the higher one—and these two energy levels correspond to distinct unit configurations related by performing a spin operation $\mathcal{M}$ on one of the spins, this key feature allows us to express the weight of the energy unit $e^{-\varepsilon_{ij}}$ as follows: 
\begin{align}
    e^{-\varepsilon_{ij}} &= e^{-\varepsilon_0} \delta_{\varepsilon_{ij}, \varepsilon_0} + e^{-\varepsilon_1} (1 - \delta_{\varepsilon_{ij}, \varepsilon_0}) \nonumber \\
    &= e^{-\varepsilon_0} \left[ P_b \delta_{\varepsilon_{ij}, \varepsilon_0} + (1 - P_b) \right],
\end{align}
where $\delta_{\varepsilon_{ij}, \varepsilon_0}$ equals 1 only if $\varepsilon_{ij} = \varepsilon_0$, and 0 otherwise. Additionally, $P_b = 1 - e^{-(\varepsilon_1 - \varepsilon_0)}$ is the bond probability. This expression can be interpreted as follows: if $\varepsilon_{ij} \neq \varepsilon_0$, only the $1 - P_b$ term remains, meaning the bond is skipped; if $\varepsilon_{ij} = \varepsilon_0$, a bond is placed with probability $P_b$ and skipped with $1 - P_b$. Furthermore, this bond placement process can be summarized by placing a bond with probability $\max[0, P_b]$, where $\varepsilon_0$ is replaced by $\varepsilon_{ij}$, and $\varepsilon_1$ is the energy resulting from applying the operation $\mathcal{M}$ to one of the spins. Through this process, cluster formation is completed. Subsequently, a spin operation $\mathcal{M}$ is applied to the spins of each cluster with a probability of $1/2$, resulting in a new configuration.

(i) For the ferromagnetic coupling model, the interlayer energy between sites is given by $-K \cos(\theta_{i,a} - \theta_{i,b})$, resulting in continuous energy levels. To achieve two discrete energy levels, we restrict spin operations to only allow flipping $\mathcal{M}: \theta \rightarrow -\theta$. Consequently, the two energy levels are defined as $\varepsilon_0 = -K \cos(\theta_{i,a} - \theta_{i,b})$ and $\varepsilon_1 = -K \cos(\theta_{i,a} + \theta_{i,b})$. Therefore, the bond probability is $P_b = 1 - e^{-2K \sin{\theta_{i,a}} \sin{\theta_{i,b}}}$. The intralayer case follows a similar approach, with the only difference being that the coupling strength is replaced by $J$. Additionally, in this constrained case, all spins are rotated by a random angle after each update is completed to ensure ergodicity.

(ii) For the {paired-phase gradient coupling} model, the term, $\cos(\theta_{i,a} + \theta_{i,b} - \theta_{j,a} - \theta_{j,b})$, requires us to consider the states of four sites simultaneously. Here, we propose three methods to reduce its energy levels to two. The main idea is to either change the spin configuration of only one layer while keeping the other layer fixed, i.e., (I); or to impose constraints on the spins at corresponding positions $\theta_{i,a}$ and $\theta_{i,b}$ in both layers and then only consider the spins of one layer, i.e., (II) and (III).

(I) \textit{Keep layer $a$ or $b$ unchanged}: Here, we keep layer $ a $ unchanged as an example. Due to this constraint, the energy contributions from layer $a$ are canceled by the difference $\varepsilon_{0} - \varepsilon_{1}$. By applying the operation $\mathcal{M}^{I}: \theta_{j,b} \rightarrow \theta_{j,b} + \pi $ to one of the two spins in the energy unit of layer $b$, we can obtain two energy levels as follows, ignoring the energy from layer $a$.
The lower one can be written as:
\begin{align}
\varepsilon_0 = -J \cos (\theta_{i,b} - \theta_{j,b}) - K \cos(\theta_{i,a} + \theta_{i,b} - \theta_{j,a} - \theta_{j,b}).
\end{align}
The higher one can be expressed as:
\begin{align}
\varepsilon_1 & = -J \cos (\theta_{i,b} - \theta_{j,b} - \pi) - K \cos(\theta_{i,a} + \theta_{i,b} - \theta_{j,a} - \theta_{j,b} - \pi) \nonumber\\ & = -\varepsilon_0.
\end{align}
Hence, the probability is given by
\begin{align}
    P_b^{I} = 1 - e^{2\varepsilon_0}.
\end{align}
Therefore, we only place bonds for layer $b$ with probability of $ \max[0, P_{b}^{I}]$. Then, we randomly flip the different clusters formed in this process with a probability of  1/2, i.e., $ \theta \rightarrow \theta + \pi $.

(II) \textit{Keep $ \theta_{i,a} - \theta_{i,b} $ unchanged}: To reduce the degree of freedom, we can make the spins in both layers change simultaneously, i.e., $ \theta_{i,a} - \theta_{i,b}$ 
 remains constant. Hence, when we apply the operation $\mathcal{M}^{II}: { \theta_{j,a}, \theta_{j,b} } \rightarrow { -\theta_{j,a}, -\theta_{j,b} }$, the energy levels of unit can be reduced to two as follows.
The lower one can be written as:
\begin{align}
\varepsilon_0 = -J \cos (\theta_{i,a} - \theta_{j,a}) - J \cos (\theta_{i,b} - \theta_{j,b}) \nonumber\\
- K \cos(\theta_{i,a} + \theta_{i,b} - \theta_{j,a} - \theta_{j,b}).
\end{align}
The higher one can be expressed as:
\begin{align}
\varepsilon_1 = -J \cos (\theta_{i,a} + \theta_{j,a}) - J \cos (\theta_{i,b} + \theta_{j,b}) \nonumber\\ - K \cos(\theta_{i,a} + \theta_{i,b} + \theta_{j,a} + \theta_{j,b}).
\end{align}
Therefore, the bond probability is given by
\begin{align}
    P_b^{II} = 1 - e^{ - (\varepsilon_1 - \varepsilon_0)}.
\end{align}
Based on this, we place bonds within one layer with probability $\max[0, P_b^{II}]$ and flip spins using operation $\mathcal{M}^{II}$.

(III) \textit{Keep $\theta_{i,a} + \theta_{i,b}$ unchanged}: Similarly, we can constrain the spins in both layers to change in opposite directions, i.e., $\theta_{i,a} + \theta_{i,b}$ remains constant. Hence, when we apply the operation $\mathcal{M}^{III}: { \theta_{j,a}, \theta_{j,b} } \rightarrow { \theta_{j,a} + \pi, \theta_{j,b} - \pi }$, the interlayer interaction energy is canceled by the difference $\varepsilon_0 - \varepsilon_1$. This allows us to ignore the energy from the interlayer and reduces the energy levels of the unit to two as follows.
The lower case can be written as:
\begin{align}
\varepsilon_0 = -J \cos (\theta_{i,a} - \theta_{j,a}) - J \cos (\theta_{i,b} - \theta_{j,b}).
\end{align}
The higher case can be expressed as:
\begin{align}
\varepsilon_1 & = -J \cos (\theta_{i,a} - \theta_{j,a} - \pi) - J \cos (\theta_{i,b} - \theta_{j,b} + \pi) \nonumber\\ & = -\varepsilon_0.
\end{align}
Hence, the probability is given by
\begin{align}
    P_b^{III} = 1 - e^{2\varepsilon_0}.
\end{align}
Based on this, we place bonds within one layer with probability $\max[0, P_b^{III}]$ and flip spins using operation $\mathcal{M}^{III}$.

Note that rotating all spins by a random angle after each update does not ensure ergodicity here. Therefore, we mix the Metropolis algorithm~\cite{Metropolis} into the update process.

\section{Estimation of the critical points and exponents}
In this Appendix, we provide detailed fitting procedures for estimating the critical points and the exponents $\eta_a$ and $\eta_p$. We employ an extrapolation method based on Eq. \eqref{eq:extrapolate} to determine the Berezinskii-Kosterlitz-Thouless (BKT) critical points. To extract the anomalous magnetic dimensions, we first use the ansatz from Eq. \eqref{eq:fitting_eta_pw}, which focuses on fitting the leading term exponent. In addition, we apply Eq. \eqref{eq:fitting_eta_ln}, which incorporates logarithmic corrections on top of the leading exponent, to refine the estimation of the anomalous magnetic dimensions.

As a precaution against correction-to-scaling terms that we missed including in the fitting ansatz, we impose a lower cutoff $L \ge L_{\rm min}$ on the data points admitted in the fit and systematically study the effect on the residuals $\chi^2$ value by increasing $L_{\rm min}$. In general, the preferred fit for any given ansatz corresponds to the smallest $L_{\rm min}$ for which the goodness of the fit is reasonable and for which subsequent increases in $L_{\rm min}$ do not cause the $\chi^2$ value to drop by vastly more than one unit per degree of freedom. In practice, by “reasonable” we mean that $\chi^2/\rm{DF} \approx 1$, where DF is the number of degrees of freedom. The systematic error is estimated by comparing estimates from various sensible fitting ansatz. 

\subsection{Estimate of the critical points}
To extract the critical points for the BKT phase transition, we employ the following ansatz~\cite{PhysRevB.65.184405}:
\begin{eqnarray}
    J(L) = J_c +\frac{\alpha}{(\ln L / L_0)^2},
    \label{eq:extrapolate}
\end{eqnarray}
where $J_c$ is the critical point we aim to determine, $\alpha$ and $L_0$ are fitting parameters, and $L$ is the system size. The function $J(L)$ represents the pseudocritical points,
which are obtained by selecting a specific value of $\xi/L$ and identifying the points where the $\xi/L$ curves for different system sizes intersect as $ J $ varies. These intersection points are determined through linear interpolation to calculate the mean and error, thereby defining $ J(L) $ for each system size.
As the system approaches the thermodynamic limit, i.e., $L\rightarrow\infty$, we obtain $J(\infty)=J_c$.

To ensure the robustness of the fit, we select multiple values of $\xi/L$, and in the following table, we present two of these values to demonstrate the stability of the fit. For the critical point, this method ensures a precision of at least two decimal places.

 \begin{figure}[!htbp]
    \centering
    \includegraphics[width=0.6\linewidth]{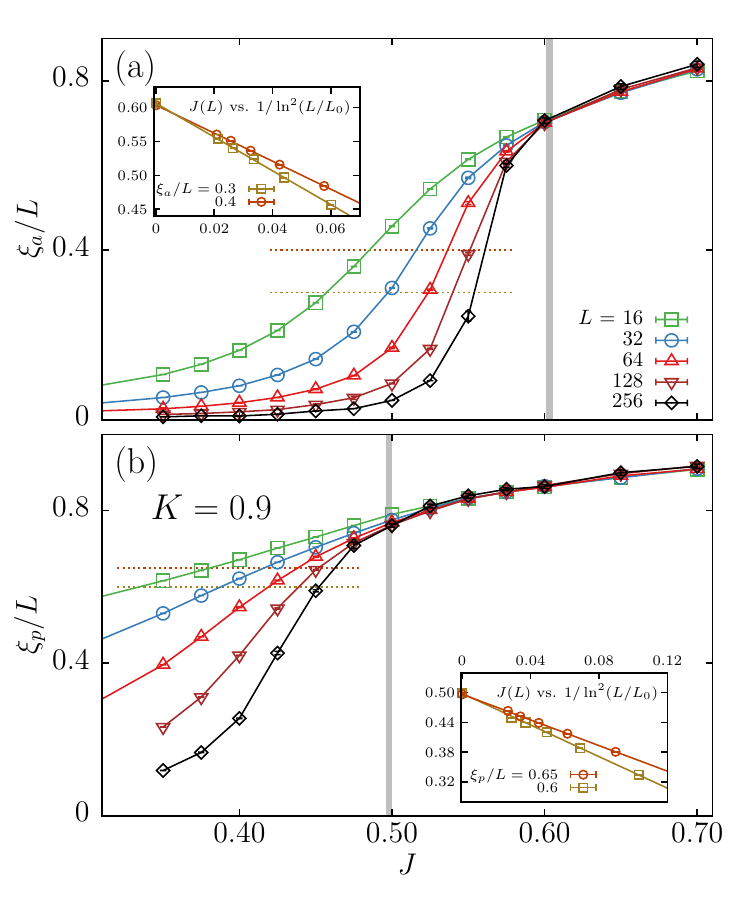}	
    \vspace*{-3mm}
    \caption{Demonstration of the estimation for the BKT critical point for the {paired-phase gradient coupling} model.} 
    \label{fig:fit_BKT}
\end{figure}

To illustrate the above process more clearly, we use the fitting procedures for the {paired-phase gradient coupling} model as an example in Fig.~\ref{fig:fit_BKT}. The dark-red and olive dashed lines in Fig.~\ref{fig:fit_BKT}(a) and (b) represent the specific values chosen for single-layer and paired spins, respectively. Through linear interpolation, we can determine the intersection points with the data curves, known as pseudocritical points, denoted as $ J(L) $. By fitting with Eq.~\eqref{eq:extrapolate}, as shown in the insets of Fig.~\ref{fig:fit_BKT}(a) and (b), and by selecting different values of $ \xi/L $, we obtain consistent results. Specifically, in the insets, as $ L \rightarrow \infty $, the two lines converge to the same intersection point. Using this method, we estimate these two models as follows.

For the ferromagnetic coupling model, we set $K = 0, 0.25, 0.50, 1.00, 2.00, 3.00$ and vary $J$ to determine the critical point $J_c$. We use the correlation length ratios for single-layer and paired spins, $\xi_a / L$ and $\xi_p / L$, respectively, to estimate the transition points. For the single-layer spin case, the fitting results are summarized in Table~\ref{tab:a_1_kc}. We observe that for each value of $K$, the estimated critical points $J_c$ are consistent within the error bars when changing $\xi_a/L$, indicating the stability of the fit. In the paired spin case, the fitting results are summarized in Table~\ref{tab:p_1_kc}. Comparing the estimated critical points from the two types of spins, we find they are consistent within the error bars, confirming that only one BKT transition occurs in this model for a fixed $K$.

\begin{table}[!ht]
    \centering
    \caption{Fitting results by using correlation length ratio of the single-layer spin  $\xi_a/L$ at various $K$ in the ferromagnetic coupling model,  using the ansatz given by Eq.~\eqref{eq:extrapolate}. }
    \vspace{20pt}
    \begin{tabular}{ll|lllllllll}
    \hline
    \hline
        $K_c$  &  $\xi_a/L$  & ~~~ $J_{c}$  & ~~ $\alpha$    &  $1/L_0$   &  $\chi^2$/DF \\ \hline
        0    & 0.30   & 1.126(6)  & -4.8(6)	 &  3.8(8)   & 3.1/3  \\
             & 0.60   & 1.116(4)   & -3.0(5)	    &  6(2)   & 8.2/3  \\
             \hline
        0.25 & 0.30 & 0.839(3) & -2.3(3) & 3.6(8) & 2.3/3	 \\
             & 0.35 & 0.840(4) & -2.4(5) & 5(2) & 8.0/3   \\
             \hline
        0.50 & 0.30 & 0.775(3) & -2.4(3) & 3.5(8) & 6.1/3  \\
             & 0.35 & 0.771(1) & -2.1(2) & 3.6(5)  & 1.1/3  \\
             \hline
        1.00 & 0.40 & 0.699(4) & -2.5(5) & 5(2)  & 3.8/3  \\
             & 0.50 & 0.698(2) & -2.3(3) & 8(2)  & 2.7/3  \\
             \hline
        2.00 & 0.40 & 0.636(3) & -2.6(3) & 6(1) & 3.8/3  \\
             & 0.50 & 0.635(2) & -2.3(2)  & 7(2)  & 1.8/3 \\
             \hline
        3.00 & 0.40 & 0.607(5) & -2.2(4)  & 4(2)  & 9.6/3 \\
             & 0.45 & 0.606(5) & -1.9(5) & 4(2)  & 8.5/3	\\ 
             \hline\hline
    \end{tabular}
    \label{tab:a_1_kc}
\end{table}

\begin{table}[!ht]
    \centering
    \caption{Fitting results by using correlation length ratio of the paired spin  $\xi_p/L$ at various $K$ in the ferromagnetic coupling model,  using the ansatz given by Eq.~\eqref{eq:extrapolate}.  }
    \vspace{20pt}
    \begin{tabular}{ll|lllllllll}
    \hline
    \hline
        $K_c$  &  $\xi_p/L$  &  ~~~$J_{c}$  & ~~ $\alpha$    & $1/L_0$    &  $\chi^2$/DF \\ \hline
        0    & 0.40   & 1.121(2)  & -3.5(3)	 &  11(2)   & 4.4/3  \\
             & 0.50   & 1.119(4)  &  -0.6(2)	&  3(2)   & 2.2/3   \\
             \hline
        0.25 & 0.20 & 0.830(1) & -0.99(5) & 1.9(1) & 1.7/3	 \\
             & 0.30 & 0.831(1) & -0.16(2) & 0.66(8) &  2.0/3   \\
             \hline
        0.50 & 0.20 & 0.767(1) & -1.38(4) & 3.3(1) & 1.8/3  \\
             & 0.25 & 0.765(1) & -0.83(5) & 2.7(3)  & 1.9/3  \\
             \hline
        1.00 & 0.20 & 0.694(1) & -1.53(9) & 3.8(4)  & 4.0/3  \\
             & 0.25 & 0.691(1) & -0.9(1) & 3.3(7)  & 3.8/3  \\
             \hline
        2.00 & 0.20 & 0.633(1) & -1.8(1) & 5.4(6) & 5.7/3  \\
             & 0.25 & 0.632(1) & -1.4(2)  & 7(2)  & 3.4/3 \\
             \hline
        3.00 & 0.20 & 0.604(4) & -1.5(3)  & 4(1)  & 7.9/3 \\
             & 0.25 & 0.605(2) & -1.2(2) & 6(2)  & 7.3/3	\\ 
             \hline\hline
    \end{tabular}
    \label{tab:p_1_kc}
\end{table}

For the {paired-phase gradient coupling} model, note that when $ K = 0 $, it is identical to the previous model. Therefore, we set $ K = 0.20, 0.50, 0.75, 0.90, 2.00, 3.00 $ to estimate the critical point $ J_c $. For the single-layer spin case, the fitting results are summarized in Table~\ref{tab:a_2_kc}. However, for the paired spin case, no phase transition is observed for $ K = 2.00, 3.00 $, and the critical points gradually deviate from those in the single-layer case as $ K $ increases, as shown in Table~\ref{tab:p_2_kc}. Therefore, we set $ J = 0, 0.30 $ and vary $ K $ to estimate the critical point $ K_c $, as shown in Table~\ref{tab:p_2_kc_}. These observations indicate the existence of a paired phase in this model.

\begin{table}[!ht]
    \centering
    \caption{Fitting results by using correlation length ratio of the paired spin  $\xi_a/L$ at various $K$ in the {paired-phase gradient coupling} model,  using the ansatz given by Eq.~\eqref{eq:extrapolate}.  }
    \vspace{20pt}
    \begin{tabular}{ll|lllllllll}
    \hline
    \hline
        $K_c$  &  $\xi_a/L$  &  ~~~$J_{c}$  & ~~ $\alpha$    & $1/L_0$   &  $\chi^2$/DF \\ \hline
        0.20 & 0.30   & 0.986(6)  & -5.4(5)	 &  5.7(9)   & 1.9/2  \\
             & 0.40   & 0.981(2)  &  -4.8(2)	&  6.8(4)   & 0.5/2   \\
             \hline
        0.50 & 0.35 & 0.741(2) & -1.9(1) & 2.1(2) & 1.3/2	 \\
             & 0.40 & 0.742(5) & -1.9(3) & 2.5(5) & 3.9/2   \\
             \hline
        0.75 & 0.35 & 0.622(2) & -1.9(1) & 2.9(3) & 1.1/2  \\
             & 0.40 & 0.624(4) & -1.9(3) & 3.5(8)  & 3.2/2  \\
             \hline
        0.90 & 0.35 & 0.606(2) & -2.4(2) & 4.1(4)  & 2.7/2  \\
             & 0.40 & 0.608(1) & -2.42(9) & 5.3(3)  & 3.1/2  \\
             \hline
        2.00 & 0.35 & 0.575(5) & -2.0(4) & 4(1) & 4.1/2  \\
             & 0.40 & 0.569(8) & -1.5(5)  & 3(1)  & 4.2/2 \\
             \hline
        3.00 & 0.40 & 0.56(1) & -1.7(8)  & 3(2)  & 3.4/1 \\
             & 0.45 & 0.574(8) & -2.3(7) & 6(3)  & 3.6/2	\\ 
             \hline\hline
    \end{tabular}
    \label{tab:a_2_kc}
\end{table}

\begin{table}[!ht]
    \centering
    \caption{Fitting results by using correlation length ratio of the paired spin  $\xi_p/L$ at various $K$ in the {paired-phase gradient coupling} model,  using the ansatz given by Eq.~\eqref{eq:extrapolate}.  }
    \vspace{20pt}
    \begin{tabular}{ll|lllllllll}
    \hline
    \hline
        $K_c$  &  $\xi_p/L$  &  ~~~$J_{c}$  & ~~ $\alpha$    & $1/L_0$    &  $\chi^2$/DF \\ \hline
        0.20 & 0.30   & 0.968(4)  & -3.4(3)	 &  4.7(7)   & 3.1/2  \\
             & 0.40   & 0.970(4)  &  -3.0(3)	&  6(1)   & 1.3/2   \\
             \hline
        0.50 & 0.30 & 0.735(2) & -1.28(9) & 1.13(9) & 2.2/2	 \\
             & 0.35 & 0.738(4) & -1.3(2) & 1.4(2) &  4.9/2   \\
             \hline
        0.75 & 0.35 & 0.596(2) & -1.24(4) & 0.69(2) & 3.7/2  \\
             & 0.40 & 0.595(2) & -1.21(6) & 0.82(4)  & 4.6/2  \\
             \hline
        0.90 & 0.60 & 0.500(1) & -1.73(8) & 1.58(1)  &  0.1/2  \\
             & 0.65 & 0.498(2) & -1.5(1) &  2.2(2) & 4.2/2  \\    
             \hline\hline
    \end{tabular}
    \label{tab:p_2_kc}
\end{table}

\begin{table}[!ht]
    \centering
    \caption{Fitting results by using correlation length ratio of the paired spin  $\xi_p/L$ at various $J$ in the {paired-phase gradient coupling} model,  using the ansatz given by Eq.~\eqref{eq:extrapolate}.  }
    \vspace{20pt}
    \begin{tabular}{ll|lllllllll}
    \hline
    \hline
        $J_c$  &  $\xi_p/L$  &  ~~~$K_{c}$  & ~~ $\alpha$    & $1/L_0$    &  $\chi^2$/DF \\ \hline
        0 & 0.60 & 1.126(3) & -4.3(4)  & 13(3)  & 2.3/2 \\
        0 & 0.70 & 1.117(5) & -1.9(5) & 8(4)  & 1.8/2	\\ 
        \hline
        0.30 & 0.50 & 1.074(6) & -5.1(6) & 9(2) & 4.5/2  \\
        0.30 & 0.60 & 1.068(8) & -4.2(9)  & 12(5)  & 2.3/2 \\
             \hline\hline
    \end{tabular}
    \label{tab:p_2_kc_}
\end{table}

\subsection{Estimate of the anomalous magnetic dimensions $\eta_a$ and $\eta_p$}

To extract the anomalous magnetic dimensions $ \eta_a $ and $ \eta_p $, we employ the ansatz presented in Eq.~\eqref{eq:fitting_eta_pw} and Eq.~\eqref{eq:fitting_eta_ln} to fit the susceptibilities for single-layer spin $ \chi_a = L^2 \langle M_a^2 \rangle $ and paired spin $ \chi_p = L^2 \langle M_p^2 \rangle $ at the BKT critical point.

By considering the correlation function scaling as $ g(r) \sim r^{-\eta} $ at the critical point, we obtain the corresponding finite-size scaling ansatz:
\begin{align}
    \chi = L^{2-\eta}(a_0 + b_1 L^{-\omega}) + c,
    \label{eq:fitting_eta_pw}
\end{align}
where $ a_0 $ and $ b_1 $ are fitting parameters, $ L^{-\omega} $ represents the finite-size correction term, and $ c $ arises from the analytic part of the free energy.

Furthermore, by incorporating the logarithmic correction term, where the correlation function scales as $ g(r) \sim r^{-\eta} (\ln r)^{-2\hat{\eta}}$ \cite{JMKosterlitz_1972}, the ansatz can be expressed as~\cite{PhysRevB.37.5986, PhysRevB.55.3580, KENNA1997583}:
\begin{align}
    \chi = L^{2-\eta}(\ln L + C_1)^{-2\hat{\eta}}(a_0 + b_1 L^{-\omega}) + c,
    \label{eq:fitting_eta_ln}    
\end{align}
where $ C_1 $ is a non-universal constant and $ \hat{\eta} $ is the correction exponent.

For the ferromagnetic coupling model, we initially leave all the fitting parameters free, but this yields unstable results. Next, by fixing $ \omega = 1 $, we obtain stable results for $ \eta $, but the error for $ c $ is large. Therefore, we fix $ c = 0 $ and obtain stable results. Subsequently, we test different values of $ \omega $ (namely, $ \omega = 0.5, 2, 3 $) to check the stability, and find that the results for $ \eta $ remain the same. As a result, we fix $c = 0$ and $ \omega = 1$ for subsequent fits.

The fitting results for $ \eta_a $ and $ \eta_p $ are summarized in Table~\ref{tab:a_1_eta} and Table~\ref{tab:p_1_eta}, respectively. Since the logarithmic corrections is not considered here, the estimated value of $ \eta_a $ is smaller than the expected standard BKT anomalous magnetic dimension of $ 1/4 $. However, we can still approximate the relation derived in Section~\ref{sec:derive}, namely $ \eta_p = 2\eta_a $ for $ K = 0 $ and $ \eta_p = 4\eta_a $ for $ K > 0 $.

\begin{table}[!ht]
    \centering
    \caption{Fitting results for the anomalous magnetic dimension $\eta_a$ of the single-layer spin at critical points $(K_c, J_c)$ in the ferromagnetic coupling model,  using the ansatz given by Eq.~\eqref{eq:fitting_eta_pw}. }
    \vspace{20pt}
    \begin{tabular}{ll|llllllllll}
    \hline
    \hline
        $K_c$  &  $J_c$  &  $L_{\rm min}$  &  $\eta_a$     &  $a_0$  &  $b_1$   &  $\chi^2$/DF \\ \hline
        0   & 1.119 & 16 & 0.2382(7)  	 & 1.011(4) &  -0.2(1)   & 8.1/6  \\
            & & 32 & 0.239(1)   	   & 1.017(7) &  -0.4(3)   & 4.4/4   \\
        \hline
        0.25 & 0.832 & 16 & 0.2334(5)  & 0.745(2) & -0.20(7) & 0.7/2	 \\
             & & 32 & 0.234(1)  & 0.747(5) & -0.3(2) & 0.5/1   \\
        \hline             
        0.50 & 0.768 & 16 & 0.2313(3)  & 0.778(1) & - & 3.5/3  \\
             & & 32 & 0.2314(4)  & 0.779(2) & - & 3.1/2  \\
        \hline
        1.00 & 0.696 & 16 & 0.2325(4)  & 0.840(2) & -0.26(6) & 0.6/2  \\
             & & 32 & 0.2333(8)  & 0.844(4) & -0.4(1) & 0.3/1  \\
        \hline
        2.00 & 0.635 & 16 & 0.2293(1)  & 0.8892(8) & 0.050(9) & 0.7/3  \\
             & & 32 & 0.2294(3)   & 0.890(1) & 0.04(2) & 0.7/2 \\
        \hline
        3.00 & 0.610 & 16 & 0.2306(3)   & 0.922(2) & -0.00(2) & 3.4/3 \\
             & & 32 & 0.2313(3)  & 0.926(1) & -0.08(3) & 0.8/2	\\ 
             \hline\hline
    \end{tabular}
    \label{tab:a_1_eta}
\end{table}

\begin{table}[!ht]
    \centering
    \caption{Fitting results for the anomalous magnetic dimension $\eta_p$ of the paired spin at critical points $(K_c, J_c)$ in the ferromagnetic coupling model,  using the ansatz given by Eq.~\eqref{eq:fitting_eta_pw}. }
    \vspace{20pt}
    \begin{tabular}{ll|lllllllllll}
    \hline
    \hline
        $K_c$  &  $J_c$  &  $L_{\rm min}$  &  $\eta_p$    &  $a_0$  &  $b_1$   &  $\chi^2$/DF \\ \hline
        0    & 1.119 & 16 & 0.4744(6) & 1.021(3) & -0.11(4) & 15.1/7  \\
             & & 32 & 0.4758(9) & 1.030(5) & -0.27(9) & 8.2/5   \\
        \hline
        0.25 & 0.832 & 16 & 0.916(2) & 1.47(1) & -1.1(2) & 9.5/3	 \\
             & & 32 & 0.920(1) & 1.51(1) & -1.9(2) & 1.9/2   \\
        \hline
        0.50 & 0.768 & 16 & 0.915(1) & 1.35(1) & -0.8(1) & 5.8/3  \\
             & & 32 & 0.918(2) & 1.37(1) & -1.2(3) & 2.8/2  \\
        \hline
        1.00 & 0.696 & 16 & 0.913(1) & 1.24(1) & -0.4(1) & 6.3/3  \\
             & & 32 & 0.916(2) & 1.27(1) & -1.0(2) & 2.1/2  \\
        \hline
        2.00 & 0.635 & 16 & 0.911(1) & 1.209(7) & -0.48(8) & 2.9/3  \\
             & & 32 & 0.911(2) & 1.21(1) & -0.4(2) & 2.8/2 \\
        \hline
        3.00 & 0.610 & 16 & 0.915(1) & 1.21(1) & -0.6(1) & 5.2/3 \\
             & &  32 & 0.918(1) & 1.23(1) & -1.0(2) & 1.9/2	\\ 
             \hline\hline
    \end{tabular}
    \label{tab:p_1_eta}
\end{table}

Furthermore, we consider logarithmic corrections. However, when we allow $ \eta $ and $ \hat{\eta} $ to vary freely, we do not obtain good fitting results. Therefore, we impose the constraint $ \hat{\eta} = -\eta / 4 $, which yields stable results.

The results are summarized in Table~\ref{tab:a_1_eta_log} and Table~\ref{tab:p_1_eta_log} for $ \eta_a $ and $ \eta_p $, respectively. We observe that the expected results are achieved, with $ \eta_a \approx \frac{1}{4} $, $ \eta_p \approx \frac{1}{2} $ for $ K = 0 $, and $ \eta_p \approx 1 $ for $ K > 0 $.

\begin{table}[!ht]
    \centering
    \caption{Fitting results for the anomalous magnetic dimension $\eta_a$ of the single-layer spin at critical points $(K_c, J_c)$ in the ferromagnetic coupling model,  using the ansatz given by Eq.~\eqref{eq:fitting_eta_ln}.}
    \vspace{20pt}
    \begin{tabular}{ll|lllllllllll}
    \hline
    \hline
        $K_c$  &  $J_c$  &  $L_{\rm min}$  &  $\eta_a$  &  $C_1$    &  $a_0$  &  $b_1$   &  $\chi^2$/DF \\ \hline
        0    & 1.119 & 16 & 0.250(5) & 4(3) & 0.81(5) & 0.04(9) & 7.2/6  \\
             & & 32 & 0.252(9) & 3(5) & 0.8(1) & 0.1(2) & 4.6/4   \\
        \hline
        0.25 & 0.832 & 16 & 0.2516(5) & - & 0.661(1) & 0.57(1) & 3.4/3	 \\
             & &  32 & 0.2523(4) & - & 0.664(1) & 0.51(2) & 0.9/2   \\
        \hline
        0.50 & 0.768 & 16 & 0.2517(4) & - & 0.701(1) & 0.49(1) & 3.3/3  \\
             & & 32 & 0.2515(8) & - & 0.701(2) & 0.50(5) & 3.3/2 \\
        \hline
        1.00 & 0.696 & 16 & 0.2520(1) & - & 0.7516(6) & 0.426(7) & 0.6/3  \\
             & & 32 & 0.2521(3) & - & 0.752(1) & 0.42(2) & 0.6/2  \\
        \hline
        2.00 & 0.635 & 16 & 0.2507(5) & - & 0.807(2) & 0.43(2) & 6.8/3  \\
             & &  32 & 0.2497(2) & - & 0.8023(8) & 0.52(1) & 0.3/2 \\
        \hline
        3.00 & 0.610 & 16 & 0.2523(4) & - & 0.836(1) & 0.39(1) & 3.3/3 \\
             & & 32 & 0.2519(6) & - & 0.835(2) & 0.43(4) & 2.3/2	\\ 
             \hline\hline
    \end{tabular}
    \label{tab:a_1_eta_log}
\end{table}

\begin{table}[!ht]
    \centering
    \caption{Fitting results for the anomalous magnetic dimension $\eta_p$ of the paired spin at critical points $(K_c, J_c)$ in the ferromagnetic coupling model,  using the ansatz given by Eq.~\eqref{eq:fitting_eta_ln}.}
    \vspace{20pt}
    \begin{tabular}{ll|lllllllllll}
    \hline
    \hline
        $K_c$  &  $J_c$  &  $L_{\rm min}$  &  $\eta_p$  &  $C_1$    &  $a_0$  &  $b_1$   &  $\chi^2$/DF \\ \hline
        0    & 1.119 & 16 & 0.509(4) & 2(1) & 0.75(4) & 0.3(1) & 4.1/6  \\
             & & 32 & 0.51(1) & 1(1) & 0.79(9) & 0.5(5) & 3.8/4    \\
        \hline
        0.25 & 0.832 & 16 & 1.0019(6) & - & 1.001(3) & 1.09(3) & 0.6/3	 \\
             & & 32 & 1.002(1) & - & 1.000(6) & 1.1(1) & 0.6/2   \\
        \hline
        0.50 & 0.768 & 16 & 1.001(1) & - & 0.917(7) & 1.21(7) & 3.9/3  \\
             & & 32 & 0.999(2) & - & 0.91(1) & 1.4(2) & 2.8/2  \\
        \hline
        1.00 & 0.696 & 16 & 0.998(1) & - & 0.843(4) & 1.30(4) & 1.3/3  \\
             & & 32 & 0.998(1) & - & 0.839(7) & 1.4(1) & 1.1/2  \\
        \hline
        2.00 & 0.635 & 16 & 0.997(2) & - & 0.82(1) & 1.2(1) & 3.9/3  \\
             & & 32 & 0.9918(3) & -  & 0.801(1) & 1.64(2) & 0.1/2 \\
        \hline
        3.00 & 0.610 & 16 & 1.001(2) & - & 0.822(7) & 1.14(7) & 5.1/3 \\
             & & 32 & 0.999(3) & - & 0.81(1) & 1.3(2) & 4.1/2	\\ 
             \hline\hline
    \end{tabular}
    \label{tab:p_1_eta_log}
\end{table}

\begin{table}[!ht]
    \centering
    \caption{Fitting results for the anomalous magnetic dimension $\eta_a$ of the single-layer spin at critical points $(K_c, J_c)$ in the {paired-phase gradient coupling} model,  using the ansatz given by Eq.~\eqref{eq:fitting_eta_pw}.}
    \vspace{20pt}
    \begin{tabular}{ll|lllllllllll}
    \hline
    \hline
        $K_c$  &  $J_c$  &  $L_{\rm min}$  &  $\eta_a$    &  $a_0$  &  $b_1$   &  $\chi^2$/DF \\ \hline
        0.20  &  0.981 & 16 & 0.225(2) & 0.97(1) & 2.0(8) & 3.4/5  \\
             & & 32 & 0.216(4) & 0.93(1) & 12(4) & 1.5/4   \\
        \hline
        0.50 & 0.742 & 16 & 0.272(6) & 0.99(3) & 0.03(6) & 3.5/5	 \\
             & & 32 & 0.25(1) & 0.85(5) & 0.4(1) & 1.2/4   \\
        \hline
        0.75 & 0.626 & 16 & 0.284(8) & 0.96(4) & 0.08(8) & 4.8/5  \\
             & & 32 & 0.28(2) & 1.0(1) & 0.1(2) & 4.8/4  \\
        \hline
        0.90 & 0.603 & 16 & 0.278(6) & 0.95(3) & 0.10(6) & 3.4/5  \\
             & & 32 & 0.26(1) & 0.87(8) & 0.3(2) & 2.8/4  \\
        \hline
        2.00 & 0.569 & 16 & 0.21(2) & 0.7(1) & 0.5(1) & 7.2/5  \\
             & & 32 & 0.11(6) & 0.4(1) & 1.2(2) & 3.2/4 \\
        \hline
        3.00 & 0.562 & 16 & 0.21(2) & 0.7(1) & 0.5(2) & 7.7/5 \\
             \hline\hline
    \end{tabular}
    \label{tab:a_2_eta}
\end{table}

\begin{table}[!ht]
    \centering
    \caption{Fitting results for the anomalous magnetic dimension $\eta_p$ of the paired spin at critical points $(K_c, J_c)$ in the {paired-phase gradient coupling} model,  using the ansatz given by Eq.~\eqref{eq:fitting_eta_pw}. }
    \vspace{20pt}
    \begin{tabular}{ll|lllllllllll}
    \hline
    \hline
        $K_c$  &  $J_c$  &  $L_{\rm min}$  &  $\eta_p$    &  $a_0$  &  $b_1$   &  $\chi^2$/DF \\ \hline
        0.20  &  0.970 & 16 & 0.421(3) & 1.03(2) & -0.01(4) & 6.4/6  \\
             & & 32 & 0.411(9) & 0.96(5) & 0.2(1) & 2.8/4   \\
        \hline
        0.50 & 0.740 & 16 & 0.371(3) & 1.11(2) & -0.10(4) & 10.0/6	 \\
             & &  32 & 0.377(8) & 1.15(5) & -0.2(1) & 4.8/4    \\
        \hline
        0.75 & 0.597 & 16 & 0.268(4) & 0.99(2) & 0.02(4) & 4.4/6  \\
             & & 32 & 0.251(8) & 0.90(4) & 0.2(1) & 1.6/4 \\
        \hline
        0.90 & 0.498 & 16 & 0.247(2) & 0.99(1) & 0.08(3) & 4.0/6  \\
             & & 32 & 0.242(6) & 0.96(3) & 0.14(9) & 1.9/4   \\
        \hline
        1.071 & 0.300 & 16 & 0.2288(4) & 0.985(2) & 0.006(6) & 2.4/6  \\
             & & 32 & 0.2301(9) & 0.994(5) & -0.02(1) & 1.4/4 \\
        \hline
        1.121 & 0 & 16 & 0.2351(3) & 1.001(2) & -0.012(5) & 2.7/6  \\
             & &  32 & 0.2357(7) & 1.005(4) & -0.02(1) & 1.8/4	\\ 
             \hline\hline
    \end{tabular}
    \label{tab:p_2_eta}
\end{table}

\begin{table}[!ht]
    \centering
    \caption{Fitting results for the anomalous magnetic dimension $\eta_a$ of the single-layer spin at critical points $(K_c, J_c)$ in the {paired-phase gradient coupling} model,  using the ansatz given by Eq.~\eqref{eq:fitting_eta_ln}.}
    \vspace{20pt}
    \begin{tabular}{ll|lllllllllll}
    \hline
    \hline
        $K_c$  &  $J_c$  &  $L_{\rm min}$  &  $\eta_a$    &  $a_0$  &  $b_1$   &  $\chi^2$/DF \\ \hline
        0.20  &  0.981 & 16 & 0.254(3) & 0.91(1) & 4.5(9) & 5.2/5  \\
             & & 32 & 0.241(5) & 0.87(1) & 16(4) & 1.8/4    \\
        \hline
        0.50 & 0.742 & 16 & 0.304(4) & 0.91(1) & 0.44(9) & 4.3/5	 \\
             & & 32 & 0.285(6) & 0.84(2) & 1.2(2) & 1.1/4    \\
        \hline
        0.75 & 0.626 & 16 & 0.318(5) & 0.88(1) & 0.5(1) & 5.2/5  \\
             & & 32 & 0.32(1) & 0.87(4) & 0.6(4) & 5.2/4  \\
        \hline
        0.90 & 0.603 & 16 & 0.312(4) & 0.88(1) & 0.54(9) & 4.1/5   \\
             & & 32 & 0.302(9) & 0.84(3) & 1.0(3) & 2.9/4  \\
        \hline
        2.00 & 0.569 & 16 & 0.27(2) & 0.80(4) & 1.0(2) & 9.0/5  \\
             & & 32 & 0.22(2) & 0.66(7) & 2.6(6) & 4.2/4 \\
        \hline
        3.00 & 0.562 & 16 & 0.26(1) & 0.81(5) & 1.0(3) & 9.3/5 \\
             & & 32 & 0.20(2) & 0.62(6) & 3.0(5) & 2.8/4 \\
             \hline\hline
    \end{tabular}
    \label{tab:a_2_eta_log}
\end{table}

\begin{table}[!ht]
    \centering
    \caption{Fitting results for the anomalous magnetic dimension $\eta_p$ of the paired spin at critical points $(K_c, J_c)$ in the {paired-phase gradient coupling} model,  using the ansatz given by Eq.~\eqref{eq:fitting_eta_ln}. }
    \vspace{20pt}
    \begin{tabular}{ll|lllllllllll}
    \hline
    \hline
        $K_c$  &  $J_c$  &  $L_{\rm min}$  &  $\eta_p$    &  $a_0$  &  $b_1$   &  $\chi^2$/DF \\ \hline
        0.20 & 0.970 & 16 & 0.445(4) & 0.78(1) & 0.38(3) & 6.7/6  \\
             &  & 32 & 0.43(1) & 0.73(4) & 0.49(9) & 2.7/4    \\
        \hline
        0.50 & 0.740 & 16 & 0.394(4) & 0.88(1) & 0.27(3) & 10.9/6	 \\
             & &  32 & 0.402(9) & 0.91(4) & 0.2(1) & 4.9/4    \\
        \hline
        0.75 & 0.597 & 16 & 0.283(4) & 0.83(1) & 0.27(3) & 4.0/6  \\
             & & 32 & 0.27(1) & 0.76(3) & 0.44(9) & 1.6/4 \\
        \hline
        0.90 & 0.498 & 16 & 0.261(3) & 0.84(1) & 0.31(2) & 6.5/6  \\
             & &  32 & 0.257(7) & 0.83(3) & 0.34(7) & 3.4/4   \\
        \hline
        1.071 & 0.300 & 16 & 0.2431(6) & 0.853(2) & 0.216(5) & 10.8/6  \\
             & & 32 & 0.245(1) & 0.861(4) & 0.19(1) & 5.3/4  \\
        \hline
        1.121 & 0 & 16 & 0.2498(5) & 0.864(2) & 0.207(5) & 6.2/6  \\
             &  & 32 & 0.2508(8) & 0.868(3) & 0.19(1) & 3.2/4	\\ 
             \hline\hline
    \end{tabular}
    \label{tab:p_2_eta_log}
\end{table}

For the {paired-phase gradient coupling} model, we follow the same procedures and summarize the fitting results in Table~\ref{tab:a_2_eta} and Table~\ref{tab:p_2_eta} for single-layer spins and paired spins, respectively. Moreover, we consider the logarithmic correction and present the fitting results in Table~\ref{tab:a_2_eta_log} and Table~\ref{tab:p_2_eta_log}. Notably, for single-layer spins, the anomalous magnetic dimension $ \eta_a $ is around $ \frac{1}{4} $; however, due to precision issues at the critical point, deviations from $ \frac{1}{4} $ occur. In contrast, for paired spins, the anomalous magnetic dimension $ \eta_p $ decreases with increasing $ K $ from $ \frac{1}{2} $ to $ \frac{1}{4} $, or increases with increasing $ J $ from $ \frac{1}{4} $ to $ \frac{1}{2} $.

\twocolumngrid
\bibliography{ref}

\begin{thebibliography}{59}%
\makeatletter
\providecommand \@ifxundefined [1]{%
 \@ifx{#1\undefined}
}%
\providecommand \@ifnum [1]{%
 \ifnum #1\expandafter \@firstoftwo
 \else \expandafter \@secondoftwo
 \fi
}%
\providecommand \@ifx [1]{%
 \ifx #1\expandafter \@firstoftwo
 \else \expandafter \@secondoftwo
 \fi
}%
\providecommand \natexlab [1]{#1}%
\providecommand \enquote  [1]{``#1''}%
\providecommand \bibnamefont  [1]{#1}%
\providecommand \bibfnamefont [1]{#1}%
\providecommand \citenamefont [1]{#1}%
\providecommand \href@noop [0]{\@secondoftwo}%
\providecommand \href [0]{\begingroup \@sanitize@url \@href}%
\providecommand \@href[1]{\@@startlink{#1}\@@href}%
\providecommand \@@href[1]{\endgroup#1\@@endlink}%
\providecommand \@sanitize@url [0]{\catcode `\\12\catcode `\$12\catcode
  `\&12\catcode `\#12\catcode `\^12\catcode `\_12\catcode `\%12\relax}%
\providecommand \@@startlink[1]{}%
\providecommand \@@endlink[0]{}%
\providecommand \url  [0]{\begingroup\@sanitize@url \@url }%
\providecommand \@url [1]{\endgroup\@href {#1}{\urlprefix }}%
\providecommand \urlprefix  [0]{URL }%
\providecommand \Eprint [0]{\href }%
\providecommand \doibase [0]{https://doi.org/}%
\providecommand \selectlanguage [0]{\@gobble}%
\providecommand \bibinfo  [0]{\@secondoftwo}%
\providecommand \bibfield  [0]{\@secondoftwo}%
\providecommand \translation [1]{[#1]}%
\providecommand \BibitemOpen [0]{}%
\providecommand \bibitemStop [0]{}%
\providecommand \bibitemNoStop [0]{.\EOS\space}%
\providecommand \EOS [0]{\spacefactor3000\relax}%
\providecommand \BibitemShut  [1]{\csname bibitem#1\endcsname}%
\let\auto@bib@innerbib\@empty
\bibitem [{\citenamefont {Cao}\ \emph {et~al.}(2018)\citenamefont {Cao},
  \citenamefont {Fatemi}, \citenamefont {Fang}, \citenamefont {Watanabe},
  \citenamefont {Taniguchi}, \citenamefont {Kaxiras},\ and\ \citenamefont
  {Jarillo-Herrero}}]{cao2018unconventional}%
  \BibitemOpen
  \bibfield  {author} {\bibinfo {author} {\bibfnamefont {Y.}~\bibnamefont
  {Cao}}, \bibinfo {author} {\bibfnamefont {V.}~\bibnamefont {Fatemi}},
  \bibinfo {author} {\bibfnamefont {S.}~\bibnamefont {Fang}}, \bibinfo {author}
  {\bibfnamefont {K.}~\bibnamefont {Watanabe}}, \bibinfo {author}
  {\bibfnamefont {T.}~\bibnamefont {Taniguchi}}, \bibinfo {author}
  {\bibfnamefont {E.}~\bibnamefont {Kaxiras}},\ and\ \bibinfo {author}
  {\bibfnamefont {P.}~\bibnamefont {Jarillo-Herrero}},\ }\bibfield  {title}
  {\bibinfo {title} {Unconventional superconductivity in magic-angle graphene
  superlattices},\ }\href {https://doi.org/https://doi.org/10.1038/nature26160}
  {\bibfield  {journal} {\bibinfo  {journal} {Nature}\ }\textbf {\bibinfo
  {volume} {556}},\ \bibinfo {pages} {43} (\bibinfo {year} {2018})}\BibitemShut
  {NoStop}%
\bibitem [{\citenamefont {Andrei}\ and\ \citenamefont
  {MacDonald}(2020)}]{andrei2020graphene}%
  \BibitemOpen
  \bibfield  {author} {\bibinfo {author} {\bibfnamefont {E.~Y.}\ \bibnamefont
  {Andrei}}\ and\ \bibinfo {author} {\bibfnamefont {A.~H.}\ \bibnamefont
  {MacDonald}},\ }\bibfield  {title} {\bibinfo {title} {Graphene bilayers with
  a twist},\ }\href
  {https://doi.org/https://doi.org/10.1038/s41563-020-00840-0} {\bibfield
  {journal} {\bibinfo  {journal} {Nat. Mater.}\ }\textbf {\bibinfo {volume}
  {19}},\ \bibinfo {pages} {1265} (\bibinfo {year} {2020})}\BibitemShut
  {NoStop}%
\bibitem [{\citenamefont {Serlin}\ \emph {et~al.}(2020)\citenamefont {Serlin},
  \citenamefont {Tschirhart}, \citenamefont {Polshyn}, \citenamefont {Zhang},
  \citenamefont {Zhu}, \citenamefont {Watanabe}, \citenamefont {Taniguchi},
  \citenamefont {Balents},\ and\ \citenamefont
  {Young}}]{doi:10.1126/science.aay5533}%
  \BibitemOpen
  \bibfield  {author} {\bibinfo {author} {\bibfnamefont {M.}~\bibnamefont
  {Serlin}}, \bibinfo {author} {\bibfnamefont {C.~L.}\ \bibnamefont
  {Tschirhart}}, \bibinfo {author} {\bibfnamefont {H.}~\bibnamefont {Polshyn}},
  \bibinfo {author} {\bibfnamefont {Y.}~\bibnamefont {Zhang}}, \bibinfo
  {author} {\bibfnamefont {J.}~\bibnamefont {Zhu}}, \bibinfo {author}
  {\bibfnamefont {K.}~\bibnamefont {Watanabe}}, \bibinfo {author}
  {\bibfnamefont {T.}~\bibnamefont {Taniguchi}}, \bibinfo {author}
  {\bibfnamefont {L.}~\bibnamefont {Balents}},\ and\ \bibinfo {author}
  {\bibfnamefont {A.~F.}\ \bibnamefont {Young}},\ }\bibfield  {title} {\bibinfo
  {title} {Intrinsic quantized anomalous hall effect in a moiré
  heterostructure},\ }\href {https://doi.org/10.1126/science.aay5533}
  {\bibfield  {journal} {\bibinfo  {journal} {Science}\ }\textbf {\bibinfo
  {volume} {367}},\ \bibinfo {pages} {900} (\bibinfo {year}
  {2020})}\BibitemShut {NoStop}%
\bibitem [{\citenamefont {Li}\ \emph {et~al.}(2021)\citenamefont {Li},
  \citenamefont {Jiang}, \citenamefont {Shen}, \citenamefont {Zhang},
  \citenamefont {Li}, \citenamefont {Tao}, \citenamefont {Devakul},
  \citenamefont {Watanabe}, \citenamefont {Taniguchi}, \citenamefont {Fu} \emph
  {et~al.}}]{li2021quantum}%
  \BibitemOpen
  \bibfield  {author} {\bibinfo {author} {\bibfnamefont {T.}~\bibnamefont
  {Li}}, \bibinfo {author} {\bibfnamefont {S.}~\bibnamefont {Jiang}}, \bibinfo
  {author} {\bibfnamefont {B.}~\bibnamefont {Shen}}, \bibinfo {author}
  {\bibfnamefont {Y.}~\bibnamefont {Zhang}}, \bibinfo {author} {\bibfnamefont
  {L.}~\bibnamefont {Li}}, \bibinfo {author} {\bibfnamefont {Z.}~\bibnamefont
  {Tao}}, \bibinfo {author} {\bibfnamefont {T.}~\bibnamefont {Devakul}},
  \bibinfo {author} {\bibfnamefont {K.}~\bibnamefont {Watanabe}}, \bibinfo
  {author} {\bibfnamefont {T.}~\bibnamefont {Taniguchi}}, \bibinfo {author}
  {\bibfnamefont {L.}~\bibnamefont {Fu}}, \emph {et~al.},\ }\bibfield  {title}
  {\bibinfo {title} {Quantum anomalous hall effect from intertwined moir{\'e}
  bands},\ }\href {https://doi.org/https://doi.org/10.1038/s41586-021-04171-1}
  {\bibfield  {journal} {\bibinfo  {journal} {Nature}\ }\textbf {\bibinfo
  {volume} {600}},\ \bibinfo {pages} {641} (\bibinfo {year}
  {2021})}\BibitemShut {NoStop}%
\bibitem [{\citenamefont {Xie}\ \emph {et~al.}(2022)\citenamefont {Xie},
  \citenamefont {Zhang}, \citenamefont {Hu}, \citenamefont {Mak},\ and\
  \citenamefont {Law}}]{PhysRevLett.128.026402}%
  \BibitemOpen
  \bibfield  {author} {\bibinfo {author} {\bibfnamefont {Y.-M.}\ \bibnamefont
  {Xie}}, \bibinfo {author} {\bibfnamefont {C.-P.}\ \bibnamefont {Zhang}},
  \bibinfo {author} {\bibfnamefont {J.-X.}\ \bibnamefont {Hu}}, \bibinfo
  {author} {\bibfnamefont {K.~F.}\ \bibnamefont {Mak}},\ and\ \bibinfo {author}
  {\bibfnamefont {K.~T.}\ \bibnamefont {Law}},\ }\bibfield  {title} {\bibinfo
  {title} {Valley-polarized quantum anomalous hall state in moir\'e
  ${\mathrm{mote}}_{2}/{\mathrm{wse}}_{2}$ heterobilayers},\ }\href
  {https://doi.org/10.1103/PhysRevLett.128.026402} {\bibfield  {journal}
  {\bibinfo  {journal} {Phys. Rev. Lett.}\ }\textbf {\bibinfo {volume} {128}},\
  \bibinfo {pages} {026402} (\bibinfo {year} {2022})}\BibitemShut {NoStop}%
\bibitem [{\citenamefont {Zhao}\ \emph {et~al.}(2024)\citenamefont {Zhao},
  \citenamefont {Kang}, \citenamefont {Zhang}, \citenamefont {Kn{\"u}ppel},
  \citenamefont {Tao}, \citenamefont {Li}, \citenamefont {Tschirhart},
  \citenamefont {Redekop}, \citenamefont {Watanabe}, \citenamefont {Taniguchi}
  \emph {et~al.}}]{zhao2024realization}%
  \BibitemOpen
  \bibfield  {author} {\bibinfo {author} {\bibfnamefont {W.}~\bibnamefont
  {Zhao}}, \bibinfo {author} {\bibfnamefont {K.}~\bibnamefont {Kang}}, \bibinfo
  {author} {\bibfnamefont {Y.}~\bibnamefont {Zhang}}, \bibinfo {author}
  {\bibfnamefont {P.}~\bibnamefont {Kn{\"u}ppel}}, \bibinfo {author}
  {\bibfnamefont {Z.}~\bibnamefont {Tao}}, \bibinfo {author} {\bibfnamefont
  {L.}~\bibnamefont {Li}}, \bibinfo {author} {\bibfnamefont {C.~L.}\
  \bibnamefont {Tschirhart}}, \bibinfo {author} {\bibfnamefont
  {E.}~\bibnamefont {Redekop}}, \bibinfo {author} {\bibfnamefont
  {K.}~\bibnamefont {Watanabe}}, \bibinfo {author} {\bibfnamefont
  {T.}~\bibnamefont {Taniguchi}}, \emph {et~al.},\ }\bibfield  {title}
  {\bibinfo {title} {Realization of the haldane chern insulator in a moir{\'e}
  lattice},\ }\href
  {https://doi.org/https://doi.org/10.1038/s41567-023-02284-0} {\bibfield
  {journal} {\bibinfo  {journal} {Nat. Phys.}\ }\textbf {\bibinfo {volume}
  {20}},\ \bibinfo {pages} {275} (\bibinfo {year} {2024})}\BibitemShut
  {NoStop}%
\bibitem [{\citenamefont {Du}\ \emph {et~al.}(2024)\citenamefont {Du},
  \citenamefont {Barral}, \citenamefont {Cantara}, \citenamefont {de~Hond},
  \citenamefont {Lu},\ and\ \citenamefont {Ketterle}}]{du2024atomic}%
  \BibitemOpen
  \bibfield  {author} {\bibinfo {author} {\bibfnamefont {L.}~\bibnamefont
  {Du}}, \bibinfo {author} {\bibfnamefont {P.}~\bibnamefont {Barral}}, \bibinfo
  {author} {\bibfnamefont {M.}~\bibnamefont {Cantara}}, \bibinfo {author}
  {\bibfnamefont {J.}~\bibnamefont {de~Hond}}, \bibinfo {author} {\bibfnamefont
  {Y.-K.}\ \bibnamefont {Lu}},\ and\ \bibinfo {author} {\bibfnamefont
  {W.}~\bibnamefont {Ketterle}},\ }\bibfield  {title} {\bibinfo {title} {Atomic
  physics on a 50-nm scale: Realization of a bilayer system of dipolar atoms},\
  }\href {https://www.science.org/doi/10.1126/science.adh3023} {\bibfield
  {journal} {\bibinfo  {journal} {Science}\ }\textbf {\bibinfo {volume}
  {384}},\ \bibinfo {pages} {546} (\bibinfo {year} {2024})}\BibitemShut
  {NoStop}%
\bibitem [{\citenamefont {Meng}\ \emph {et~al.}(2023)\citenamefont {Meng},
  \citenamefont {Wang}, \citenamefont {Han}, \citenamefont {Liu}, \citenamefont
  {Wen}, \citenamefont {Gao}, \citenamefont {Wang}, \citenamefont {Chin},\ and\
  \citenamefont {Zhang}}]{meng2023atomic}%
  \BibitemOpen
  \bibfield  {author} {\bibinfo {author} {\bibfnamefont {Z.}~\bibnamefont
  {Meng}}, \bibinfo {author} {\bibfnamefont {L.}~\bibnamefont {Wang}}, \bibinfo
  {author} {\bibfnamefont {W.}~\bibnamefont {Han}}, \bibinfo {author}
  {\bibfnamefont {F.}~\bibnamefont {Liu}}, \bibinfo {author} {\bibfnamefont
  {K.}~\bibnamefont {Wen}}, \bibinfo {author} {\bibfnamefont {C.}~\bibnamefont
  {Gao}}, \bibinfo {author} {\bibfnamefont {P.}~\bibnamefont {Wang}}, \bibinfo
  {author} {\bibfnamefont {C.}~\bibnamefont {Chin}},\ and\ \bibinfo {author}
  {\bibfnamefont {J.}~\bibnamefont {Zhang}},\ }\bibfield  {title} {\bibinfo
  {title} {Atomic bose--einstein condensate in twisted-bilayer optical
  lattices},\ }\href
  {https://doi.org/https://doi.org/10.1038/s41586-023-05695-4} {\bibfield
  {journal} {\bibinfo  {journal} {Nature}\ }\textbf {\bibinfo {volume} {615}},\
  \bibinfo {pages} {231} (\bibinfo {year} {2023})}\BibitemShut {NoStop}%
\bibitem [{\citenamefont {Wang}\ \emph {et~al.}(2020)\citenamefont {Wang},
  \citenamefont {Zheng}, \citenamefont {Chen}, \citenamefont {Huang},
  \citenamefont {Kartashov}, \citenamefont {Torner}, \citenamefont {Konotop},\
  and\ \citenamefont {Ye}}]{wang2020localization}%
  \BibitemOpen
  \bibfield  {author} {\bibinfo {author} {\bibfnamefont {P.}~\bibnamefont
  {Wang}}, \bibinfo {author} {\bibfnamefont {Y.}~\bibnamefont {Zheng}},
  \bibinfo {author} {\bibfnamefont {X.}~\bibnamefont {Chen}}, \bibinfo {author}
  {\bibfnamefont {C.}~\bibnamefont {Huang}}, \bibinfo {author} {\bibfnamefont
  {Y.~V.}\ \bibnamefont {Kartashov}}, \bibinfo {author} {\bibfnamefont
  {L.}~\bibnamefont {Torner}}, \bibinfo {author} {\bibfnamefont {V.~V.}\
  \bibnamefont {Konotop}},\ and\ \bibinfo {author} {\bibfnamefont
  {F.}~\bibnamefont {Ye}},\ }\bibfield  {title} {\bibinfo {title} {Localization
  and delocalization of light in photonic moir{\'e} lattices},\ }\href
  {https://doi.org/https://doi.org/10.1038/s41586-019-1851-6} {\bibfield
  {journal} {\bibinfo  {journal} {Nature}\ }\textbf {\bibinfo {volume} {577}},\
  \bibinfo {pages} {42} (\bibinfo {year} {2020})}\BibitemShut {NoStop}%
\bibitem [{\citenamefont {Parga}\ and\ \citenamefont {{Van
  Himbergen}}(1980)}]{PARGA1980607}%
  \BibitemOpen
  \bibfield  {author} {\bibinfo {author} {\bibfnamefont {N.}~\bibnamefont
  {Parga}}\ and\ \bibinfo {author} {\bibfnamefont {J.}~\bibnamefont {{Van
  Himbergen}}},\ }\bibfield  {title} {\bibinfo {title} {Critical behavior of
  the double layer classical xy-model},\ }\href
  {https://doi.org/https://doi.org/10.1016/0038-1098(80)90592-X} {\bibfield
  {journal} {\bibinfo  {journal} {Solid State Commun.}\ }\textbf {\bibinfo
  {volume} {35}},\ \bibinfo {pages} {607} (\bibinfo {year} {1980})}\BibitemShut
  {NoStop}%
\bibitem [{\citenamefont {Jiang}\ \emph {et~al.}(1996)\citenamefont {Jiang},
  \citenamefont {Stoebe},\ and\ \citenamefont {Huang}}]{PhysRevLett.76.2910}%
  \BibitemOpen
  \bibfield  {author} {\bibinfo {author} {\bibfnamefont {I.~M.}\ \bibnamefont
  {Jiang}}, \bibinfo {author} {\bibfnamefont {T.}~\bibnamefont {Stoebe}},\ and\
  \bibinfo {author} {\bibfnamefont {C.~C.}\ \bibnamefont {Huang}},\ }\bibfield
  {title} {\bibinfo {title} {Monte carlo studies of helicity modulus and heat
  capacity of a coupled xy model in two dimensions},\ }\href
  {https://doi.org/10.1103/PhysRevLett.76.2910} {\bibfield  {journal} {\bibinfo
   {journal} {Phys. Rev. Lett.}\ }\textbf {\bibinfo {volume} {76}},\ \bibinfo
  {pages} {2910} (\bibinfo {year} {1996})}\BibitemShut {NoStop}%
\bibitem [{\citenamefont {Perali}\ \emph {et~al.}(2013)\citenamefont {Perali},
  \citenamefont {Neilson},\ and\ \citenamefont
  {Hamilton}}]{PhysRevLett.110.146803}%
  \BibitemOpen
  \bibfield  {author} {\bibinfo {author} {\bibfnamefont {A.}~\bibnamefont
  {Perali}}, \bibinfo {author} {\bibfnamefont {D.}~\bibnamefont {Neilson}},\
  and\ \bibinfo {author} {\bibfnamefont {A.~R.}\ \bibnamefont {Hamilton}},\
  }\bibfield  {title} {\bibinfo {title} {High-temperature superfluidity in
  double-bilayer graphene},\ }\href
  {https://doi.org/10.1103/PhysRevLett.110.146803} {\bibfield  {journal}
  {\bibinfo  {journal} {Phys. Rev. Lett.}\ }\textbf {\bibinfo {volume} {110}},\
  \bibinfo {pages} {146803} (\bibinfo {year} {2013})}\BibitemShut {NoStop}%
\bibitem [{\citenamefont {Babaev}(2002)}]{PhysRevLett.89.067001}%
  \BibitemOpen
  \bibfield  {author} {\bibinfo {author} {\bibfnamefont {E.}~\bibnamefont
  {Babaev}},\ }\bibfield  {title} {\bibinfo {title} {Vortices with fractional
  flux in two-gap superconductors and in extended faddeev model},\ }\href
  {https://doi.org/10.1103/PhysRevLett.89.067001} {\bibfield  {journal}
  {\bibinfo  {journal} {Phys. Rev. Lett.}\ }\textbf {\bibinfo {volume} {89}},\
  \bibinfo {pages} {067001} (\bibinfo {year} {2002})}\BibitemShut {NoStop}%
\bibitem [{\citenamefont {Babaev}(2004)}]{BABAEV2004397}%
  \BibitemOpen
  \bibfield  {author} {\bibinfo {author} {\bibfnamefont {E.}~\bibnamefont
  {Babaev}},\ }\bibfield  {title} {\bibinfo {title} {Phase diagram of planar
  u(1)×u(1) superconductor: Condensation of vortices with fractional flux and
  a superfluid state},\ }\href
  {https://doi.org/https://doi.org/10.1016/j.nuclphysb.2004.02.021} {\bibfield
  {journal} {\bibinfo  {journal} {Nucl. Phys. B.}\ }\textbf {\bibinfo {volume}
  {686}},\ \bibinfo {pages} {397} (\bibinfo {year} {2004})}\BibitemShut
  {NoStop}%
\bibitem [{\citenamefont {Bojesen}\ \emph {et~al.}(2013)\citenamefont
  {Bojesen}, \citenamefont {Babaev},\ and\ \citenamefont
  {Sudb\o{}}}]{PhysRevB.88.220511}%
  \BibitemOpen
  \bibfield  {author} {\bibinfo {author} {\bibfnamefont {T.~A.}\ \bibnamefont
  {Bojesen}}, \bibinfo {author} {\bibfnamefont {E.}~\bibnamefont {Babaev}},\
  and\ \bibinfo {author} {\bibfnamefont {A.}~\bibnamefont {Sudb\o{}}},\
  }\bibfield  {title} {\bibinfo {title} {Time reversal symmetry breakdown in
  normal and superconducting states in frustrated three-band systems},\ }\href
  {https://doi.org/10.1103/PhysRevB.88.220511} {\bibfield  {journal} {\bibinfo
  {journal} {Phys. Rev. B}\ }\textbf {\bibinfo {volume} {88}},\ \bibinfo
  {pages} {220511} (\bibinfo {year} {2013})}\BibitemShut {NoStop}%
\bibitem [{\citenamefont {Macia}\ \emph {et~al.}(2014)\citenamefont {Macia},
  \citenamefont {Astrakharchik}, \citenamefont {Mazzanti}, \citenamefont
  {Giorgini},\ and\ \citenamefont {Boronat}}]{PhysRevA.90.043623}%
  \BibitemOpen
  \bibfield  {author} {\bibinfo {author} {\bibfnamefont {A.}~\bibnamefont
  {Macia}}, \bibinfo {author} {\bibfnamefont {G.~E.}\ \bibnamefont
  {Astrakharchik}}, \bibinfo {author} {\bibfnamefont {F.}~\bibnamefont
  {Mazzanti}}, \bibinfo {author} {\bibfnamefont {S.}~\bibnamefont {Giorgini}},\
  and\ \bibinfo {author} {\bibfnamefont {J.}~\bibnamefont {Boronat}},\
  }\bibfield  {title} {\bibinfo {title} {Single-particle versus pair
  superfluidity in a bilayer system of dipolar bosons},\ }\href
  {https://doi.org/10.1103/PhysRevA.90.043623} {\bibfield  {journal} {\bibinfo
  {journal} {Phys. Rev. A}\ }\textbf {\bibinfo {volume} {90}},\ \bibinfo
  {pages} {043623} (\bibinfo {year} {2014})}\BibitemShut {NoStop}%
\bibitem [{\citenamefont {Baranov}\ \emph {et~al.}(2012)\citenamefont
  {Baranov}, \citenamefont {Dalmonte}, \citenamefont {Pupillo},\ and\
  \citenamefont {Zoller}}]{doi:10.1021/cr2003568}%
  \BibitemOpen
  \bibfield  {author} {\bibinfo {author} {\bibfnamefont {M.~A.}\ \bibnamefont
  {Baranov}}, \bibinfo {author} {\bibfnamefont {M.}~\bibnamefont {Dalmonte}},
  \bibinfo {author} {\bibfnamefont {G.}~\bibnamefont {Pupillo}},\ and\ \bibinfo
  {author} {\bibfnamefont {P.}~\bibnamefont {Zoller}},\ }\bibfield  {title}
  {\bibinfo {title} {Condensed matter theory of dipolar quantum gases},\ }\href
  {https://doi.org/10.1021/cr2003568} {\bibfield  {journal} {\bibinfo
  {journal} {Chem. Rev.}\ }\textbf {\bibinfo {volume} {112}},\ \bibinfo {pages}
  {5012} (\bibinfo {year} {2012})}\BibitemShut {NoStop}%
\bibitem [{\citenamefont {Bighin}\ \emph {et~al.}(2019)\citenamefont {Bighin},
  \citenamefont {Defenu}, \citenamefont {N\'andori}, \citenamefont
  {Salasnich},\ and\ \citenamefont {Trombettoni}}]{PhysRevLett.123.100601}%
  \BibitemOpen
  \bibfield  {author} {\bibinfo {author} {\bibfnamefont {G.}~\bibnamefont
  {Bighin}}, \bibinfo {author} {\bibfnamefont {N.}~\bibnamefont {Defenu}},
  \bibinfo {author} {\bibfnamefont {I.}~\bibnamefont {N\'andori}}, \bibinfo
  {author} {\bibfnamefont {L.}~\bibnamefont {Salasnich}},\ and\ \bibinfo
  {author} {\bibfnamefont {A.}~\bibnamefont {Trombettoni}},\ }\bibfield
  {title} {\bibinfo {title} {Berezinskii-kosterlitz-thouless paired phase in
  coupled $xy$ models},\ }\href
  {https://doi.org/10.1103/PhysRevLett.123.100601} {\bibfield  {journal}
  {\bibinfo  {journal} {Phys. Rev. Lett.}\ }\textbf {\bibinfo {volume} {123}},\
  \bibinfo {pages} {100601} (\bibinfo {year} {2019})}\BibitemShut {NoStop}%
\bibitem [{\citenamefont {Gonz\'alez-Tudela}\ and\ \citenamefont
  {Cirac}(2019)}]{PhysRevA.100.053604}%
  \BibitemOpen
  \bibfield  {author} {\bibinfo {author} {\bibfnamefont {A.}~\bibnamefont
  {Gonz\'alez-Tudela}}\ and\ \bibinfo {author} {\bibfnamefont {J.~I.}\
  \bibnamefont {Cirac}},\ }\bibfield  {title} {\bibinfo {title} {Cold atoms in
  twisted-bilayer optical potentials},\ }\href
  {https://doi.org/10.1103/PhysRevA.100.053604} {\bibfield  {journal} {\bibinfo
   {journal} {Phys. Rev. A}\ }\textbf {\bibinfo {volume} {100}},\ \bibinfo
  {pages} {053604} (\bibinfo {year} {2019})}\BibitemShut {NoStop}%
\bibitem [{\citenamefont {Song}\ and\ \citenamefont
  {Zhang}(2022)}]{PhysRevLett.128.195301}%
  \BibitemOpen
  \bibfield  {author} {\bibinfo {author} {\bibfnamefont {F.-F.}\ \bibnamefont
  {Song}}\ and\ \bibinfo {author} {\bibfnamefont {G.-M.}\ \bibnamefont
  {Zhang}},\ }\bibfield  {title} {\bibinfo {title} {Phase coherence of pairs of
  cooper pairs as quasi-long-range order of half-vortex pairs in a
  two-dimensional bilayer system},\ }\href
  {https://doi.org/10.1103/PhysRevLett.128.195301} {\bibfield  {journal}
  {\bibinfo  {journal} {Phys. Rev. Lett.}\ }\textbf {\bibinfo {volume} {128}},\
  \bibinfo {pages} {195301} (\bibinfo {year} {2022})}\BibitemShut {NoStop}%
\bibitem [{\citenamefont {Masini}\ \emph {et~al.}(2025)\citenamefont {Masini},
  \citenamefont {Cuccoli}, \citenamefont {Rettori}, \citenamefont
  {Trombettoni},\ and\ \citenamefont {Cinti}}]{PhysRevB.111.094415}%
  \BibitemOpen
  \bibfield  {author} {\bibinfo {author} {\bibfnamefont {A.}~\bibnamefont
  {Masini}}, \bibinfo {author} {\bibfnamefont {A.}~\bibnamefont {Cuccoli}},
  \bibinfo {author} {\bibfnamefont {A.}~\bibnamefont {Rettori}}, \bibinfo
  {author} {\bibfnamefont {A.}~\bibnamefont {Trombettoni}},\ and\ \bibinfo
  {author} {\bibfnamefont {F.}~\bibnamefont {Cinti}},\ }\bibfield  {title}
  {\bibinfo {title} {Helicity modulus in the bilayer xy model by the monte
  carlo worm algorithm},\ }\href {https://doi.org/10.1103/PhysRevB.111.094415}
  {\bibfield  {journal} {\bibinfo  {journal} {Phys. Rev. B}\ }\textbf {\bibinfo
  {volume} {111}},\ \bibinfo {pages} {094415} (\bibinfo {year}
  {2025})}\BibitemShut {NoStop}%
\bibitem [{\citenamefont {Tao}\ \emph {et~al.}(2024)\citenamefont {Tao},
  \citenamefont {Shen}, \citenamefont {Jiang}, \citenamefont {Li},
  \citenamefont {Li}, \citenamefont {Ma}, \citenamefont {Zhao}, \citenamefont
  {Hu}, \citenamefont {Pistunova}, \citenamefont {Watanabe}, \citenamefont
  {Taniguchi}, \citenamefont {Heinz}, \citenamefont {Mak},\ and\ \citenamefont
  {Shan}}]{PhysRevX.14.011004}%
  \BibitemOpen
  \bibfield  {author} {\bibinfo {author} {\bibfnamefont {Z.}~\bibnamefont
  {Tao}}, \bibinfo {author} {\bibfnamefont {B.}~\bibnamefont {Shen}}, \bibinfo
  {author} {\bibfnamefont {S.}~\bibnamefont {Jiang}}, \bibinfo {author}
  {\bibfnamefont {T.}~\bibnamefont {Li}}, \bibinfo {author} {\bibfnamefont
  {L.}~\bibnamefont {Li}}, \bibinfo {author} {\bibfnamefont {L.}~\bibnamefont
  {Ma}}, \bibinfo {author} {\bibfnamefont {W.}~\bibnamefont {Zhao}}, \bibinfo
  {author} {\bibfnamefont {J.}~\bibnamefont {Hu}}, \bibinfo {author}
  {\bibfnamefont {K.}~\bibnamefont {Pistunova}}, \bibinfo {author}
  {\bibfnamefont {K.}~\bibnamefont {Watanabe}}, \bibinfo {author}
  {\bibfnamefont {T.}~\bibnamefont {Taniguchi}}, \bibinfo {author}
  {\bibfnamefont {T.~F.}\ \bibnamefont {Heinz}}, \bibinfo {author}
  {\bibfnamefont {K.~F.}\ \bibnamefont {Mak}},\ and\ \bibinfo {author}
  {\bibfnamefont {J.}~\bibnamefont {Shan}},\ }\bibfield  {title} {\bibinfo
  {title} {Valley-coherent quantum anomalous hall state in ab-stacked
  ${\mathrm{mote}}_{2}/{\mathrm{w}\mathrm{s}\mathrm{e}}_{2}$ bilayers},\ }\href
  {https://doi.org/10.1103/PhysRevX.14.011004} {\bibfield  {journal} {\bibinfo
  {journal} {Phys. Rev. X}\ }\textbf {\bibinfo {volume} {14}},\ \bibinfo
  {pages} {011004} (\bibinfo {year} {2024})}\BibitemShut {NoStop}%
\bibitem [{\citenamefont {Qu}\ \emph {et~al.}(2024)\citenamefont {Qu},
  \citenamefont {Qu}, \citenamefont {Chen}, \citenamefont {Wu}, \citenamefont
  {Yang}, \citenamefont {Li},\ and\ \citenamefont
  {Su}}]{PhysRevLett.132.036502}%
  \BibitemOpen
  \bibfield  {author} {\bibinfo {author} {\bibfnamefont {X.-Z.}\ \bibnamefont
  {Qu}}, \bibinfo {author} {\bibfnamefont {D.-W.}\ \bibnamefont {Qu}}, \bibinfo
  {author} {\bibfnamefont {J.}~\bibnamefont {Chen}}, \bibinfo {author}
  {\bibfnamefont {C.}~\bibnamefont {Wu}}, \bibinfo {author} {\bibfnamefont
  {F.}~\bibnamefont {Yang}}, \bibinfo {author} {\bibfnamefont {W.}~\bibnamefont
  {Li}},\ and\ \bibinfo {author} {\bibfnamefont {G.}~\bibnamefont {Su}},\
  }\bibfield  {title} {\bibinfo {title} {Bilayer
  ${t\text{\ensuremath{-}}J\text{\ensuremath{-}}J}_{\ensuremath{\perp}}$ model
  and magnetically mediated pairing in the pressurized nickelate
  ${\mathrm{la}}_{3}{\mathrm{ni}}_{2}{\mathrm{o}}_{7}$},\ }\href
  {https://doi.org/10.1103/PhysRevLett.132.036502} {\bibfield  {journal}
  {\bibinfo  {journal} {Phys. Rev. Lett.}\ }\textbf {\bibinfo {volume} {132}},\
  \bibinfo {pages} {036502} (\bibinfo {year} {2024})}\BibitemShut {NoStop}%
\bibitem [{\citenamefont {Lin}\ \emph {et~al.}(2022)\citenamefont {Lin},
  \citenamefont {Chen}, \citenamefont {Li}, \citenamefont {Meng},\ and\
  \citenamefont {Shi}}]{PhysRevLett.128.157201}%
  \BibitemOpen
  \bibfield  {author} {\bibinfo {author} {\bibfnamefont {X.}~\bibnamefont
  {Lin}}, \bibinfo {author} {\bibfnamefont {B.-B.}\ \bibnamefont {Chen}},
  \bibinfo {author} {\bibfnamefont {W.}~\bibnamefont {Li}}, \bibinfo {author}
  {\bibfnamefont {Z.~Y.}\ \bibnamefont {Meng}},\ and\ \bibinfo {author}
  {\bibfnamefont {T.}~\bibnamefont {Shi}},\ }\bibfield  {title} {\bibinfo
  {title} {Exciton proliferation and fate of the topological mott insulator in
  a twisted bilayer graphene lattice model},\ }\href
  {https://doi.org/10.1103/PhysRevLett.128.157201} {\bibfield  {journal}
  {\bibinfo  {journal} {Phys. Rev. Lett.}\ }\textbf {\bibinfo {volume} {128}},\
  \bibinfo {pages} {157201} (\bibinfo {year} {2022})}\BibitemShut {NoStop}%
\bibitem [{\citenamefont {Zhang}\ \emph {et~al.}(2022)\citenamefont {Zhang},
  \citenamefont {Sun}, \citenamefont {Li}, \citenamefont {Pan},\ and\
  \citenamefont {Meng}}]{PhysRevB.106.184517}%
  \BibitemOpen
  \bibfield  {author} {\bibinfo {author} {\bibfnamefont {X.}~\bibnamefont
  {Zhang}}, \bibinfo {author} {\bibfnamefont {K.}~\bibnamefont {Sun}}, \bibinfo
  {author} {\bibfnamefont {H.}~\bibnamefont {Li}}, \bibinfo {author}
  {\bibfnamefont {G.}~\bibnamefont {Pan}},\ and\ \bibinfo {author}
  {\bibfnamefont {Z.~Y.}\ \bibnamefont {Meng}},\ }\bibfield  {title} {\bibinfo
  {title} {Superconductivity and bosonic fluid emerging from moir\'e flat
  bands},\ }\href {https://doi.org/10.1103/PhysRevB.106.184517} {\bibfield
  {journal} {\bibinfo  {journal} {Phys. Rev. B}\ }\textbf {\bibinfo {volume}
  {106}},\ \bibinfo {pages} {184517} (\bibinfo {year} {2022})}\BibitemShut
  {NoStop}%
\bibitem [{\citenamefont {Ma}\ \emph {et~al.}(2025)\citenamefont {Ma},
  \citenamefont {Huang}, \citenamefont {Zhang}, \citenamefont {Hu},
  \citenamefont {Zhou}, \citenamefont {Feng}, \citenamefont {Li}, \citenamefont
  {Chen}, \citenamefont {Lou}, \citenamefont {Zhang} \emph
  {et~al.}}]{ma2025magnetic}%
  \BibitemOpen
  \bibfield  {author} {\bibinfo {author} {\bibfnamefont {Y.}~\bibnamefont
  {Ma}}, \bibinfo {author} {\bibfnamefont {M.}~\bibnamefont {Huang}}, \bibinfo
  {author} {\bibfnamefont {X.}~\bibnamefont {Zhang}}, \bibinfo {author}
  {\bibfnamefont {W.}~\bibnamefont {Hu}}, \bibinfo {author} {\bibfnamefont
  {Z.}~\bibnamefont {Zhou}}, \bibinfo {author} {\bibfnamefont {K.}~\bibnamefont
  {Feng}}, \bibinfo {author} {\bibfnamefont {W.}~\bibnamefont {Li}}, \bibinfo
  {author} {\bibfnamefont {Y.}~\bibnamefont {Chen}}, \bibinfo {author}
  {\bibfnamefont {C.}~\bibnamefont {Lou}}, \bibinfo {author} {\bibfnamefont
  {W.}~\bibnamefont {Zhang}}, \emph {et~al.},\ }\bibfield  {title} {\bibinfo
  {title} {Magnetic bloch states at integer flux quanta induced by
  super-moir{\'e} potential in graphene aligned with twisted boron nitride},\
  }\href {https://doi.org/https://doi.org/10.1038/s41467-025-57111-2}
  {\bibfield  {journal} {\bibinfo  {journal} {Nat. Commun.}\ }\textbf {\bibinfo
  {volume} {16}},\ \bibinfo {pages} {1860} (\bibinfo {year}
  {2025})}\BibitemShut {NoStop}%
\bibitem [{\citenamefont {Zhang}\ \emph {et~al.}(2025)\citenamefont {Zhang},
  \citenamefont {Fan}, \citenamefont {Capogrosso-Sansone},\ and\ \citenamefont
  {Deng}}]{PhysRevB.111.024511}%
  \BibitemOpen
  \bibfield  {author} {\bibinfo {author} {\bibfnamefont {C.}~\bibnamefont
  {Zhang}}, \bibinfo {author} {\bibfnamefont {Z.}~\bibnamefont {Fan}}, \bibinfo
  {author} {\bibfnamefont {B.}~\bibnamefont {Capogrosso-Sansone}},\ and\
  \bibinfo {author} {\bibfnamefont {Y.}~\bibnamefont {Deng}},\ }\bibfield
  {title} {\bibinfo {title} {Dipolar bosons in a twisted bilayer geometry},\
  }\href {https://doi.org/10.1103/PhysRevB.111.024511} {\bibfield  {journal}
  {\bibinfo  {journal} {Phys. Rev. B}\ }\textbf {\bibinfo {volume} {111}},\
  \bibinfo {pages} {024511} (\bibinfo {year} {2025})}\BibitemShut {NoStop}%
\bibitem [{\citenamefont {Grinenko}\ \emph {et~al.}(2021)\citenamefont
  {Grinenko}, \citenamefont {Weston}, \citenamefont {Caglieris}, \citenamefont
  {Wuttke}, \citenamefont {Hess}, \citenamefont {Gottschall}, \citenamefont
  {Maccari}, \citenamefont {Gorbunov}, \citenamefont {Zherlitsyn},
  \citenamefont {Wosnitza} \emph {et~al.}}]{grinenko2021state}%
  \BibitemOpen
  \bibfield  {author} {\bibinfo {author} {\bibfnamefont {V.}~\bibnamefont
  {Grinenko}}, \bibinfo {author} {\bibfnamefont {D.}~\bibnamefont {Weston}},
  \bibinfo {author} {\bibfnamefont {F.}~\bibnamefont {Caglieris}}, \bibinfo
  {author} {\bibfnamefont {C.}~\bibnamefont {Wuttke}}, \bibinfo {author}
  {\bibfnamefont {C.}~\bibnamefont {Hess}}, \bibinfo {author} {\bibfnamefont
  {T.}~\bibnamefont {Gottschall}}, \bibinfo {author} {\bibfnamefont
  {I.}~\bibnamefont {Maccari}}, \bibinfo {author} {\bibfnamefont
  {D.}~\bibnamefont {Gorbunov}}, \bibinfo {author} {\bibfnamefont
  {S.}~\bibnamefont {Zherlitsyn}}, \bibinfo {author} {\bibfnamefont
  {J.}~\bibnamefont {Wosnitza}}, \emph {et~al.},\ }\bibfield  {title} {\bibinfo
  {title} {State with spontaneously broken time-reversal symmetry above the
  superconducting phase transition},\ }\href@noop {} {\bibfield  {journal}
  {\bibinfo  {journal} {Nature Physics}\ }\textbf {\bibinfo {volume} {17}},\
  \bibinfo {pages} {1254} (\bibinfo {year} {2021})}\BibitemShut {NoStop}%
\bibitem [{\citenamefont {Shipulin}\ \emph {et~al.}(2023)\citenamefont
  {Shipulin}, \citenamefont {Stegani}, \citenamefont {Maccari}, \citenamefont
  {Kihou}, \citenamefont {Lee}, \citenamefont {Hu}, \citenamefont {Zheng},
  \citenamefont {Yang}, \citenamefont {Li}, \citenamefont {Yim} \emph
  {et~al.}}]{shipulin2023calorimetric}%
  \BibitemOpen
  \bibfield  {author} {\bibinfo {author} {\bibfnamefont {I.}~\bibnamefont
  {Shipulin}}, \bibinfo {author} {\bibfnamefont {N.}~\bibnamefont {Stegani}},
  \bibinfo {author} {\bibfnamefont {I.}~\bibnamefont {Maccari}}, \bibinfo
  {author} {\bibfnamefont {K.}~\bibnamefont {Kihou}}, \bibinfo {author}
  {\bibfnamefont {C.-H.}\ \bibnamefont {Lee}}, \bibinfo {author} {\bibfnamefont
  {Q.}~\bibnamefont {Hu}}, \bibinfo {author} {\bibfnamefont {Y.}~\bibnamefont
  {Zheng}}, \bibinfo {author} {\bibfnamefont {F.}~\bibnamefont {Yang}},
  \bibinfo {author} {\bibfnamefont {Y.}~\bibnamefont {Li}}, \bibinfo {author}
  {\bibfnamefont {C.-M.}\ \bibnamefont {Yim}}, \emph {et~al.},\ }\bibfield
  {title} {\bibinfo {title} {Calorimetric evidence for two phase transitions in
  ba1- x k x fe2as2 with fermion pairing and quadrupling states},\ }\href@noop
  {} {\bibfield  {journal} {\bibinfo  {journal} {Nature Communications}\
  }\textbf {\bibinfo {volume} {14}},\ \bibinfo {pages} {6734} (\bibinfo {year}
  {2023})}\BibitemShut {NoStop}%
\bibitem [{\citenamefont {Halcrow}\ \emph {et~al.}(2024)\citenamefont
  {Halcrow}, \citenamefont {Shipulin}, \citenamefont {Caglieris}, \citenamefont
  {Li}, \citenamefont {Wosnitza}, \citenamefont {Klauss}, \citenamefont
  {Zherlitsyn}, \citenamefont {Grinenko},\ and\ \citenamefont
  {Babaev}}]{halcrow2024probing}%
  \BibitemOpen
  \bibfield  {author} {\bibinfo {author} {\bibfnamefont {C.}~\bibnamefont
  {Halcrow}}, \bibinfo {author} {\bibfnamefont {I.}~\bibnamefont {Shipulin}},
  \bibinfo {author} {\bibfnamefont {F.}~\bibnamefont {Caglieris}}, \bibinfo
  {author} {\bibfnamefont {Y.}~\bibnamefont {Li}}, \bibinfo {author}
  {\bibfnamefont {J.}~\bibnamefont {Wosnitza}}, \bibinfo {author}
  {\bibfnamefont {H.-H.}\ \bibnamefont {Klauss}}, \bibinfo {author}
  {\bibfnamefont {S.}~\bibnamefont {Zherlitsyn}}, \bibinfo {author}
  {\bibfnamefont {V.}~\bibnamefont {Grinenko}},\ and\ \bibinfo {author}
  {\bibfnamefont {E.}~\bibnamefont {Babaev}},\ }\bibfield  {title} {\bibinfo
  {title} {Probing electron quadrupling order through ultrasound},\ }\href@noop
  {} {\bibfield  {journal} {\bibinfo  {journal} {arXiv preprint
  arXiv:2404.03020}\ } (\bibinfo {year} {2024})}\BibitemShut {NoStop}%
\bibitem [{\citenamefont {Berezinsky}(1972)}]{berezinsky1972destruction}%
  \BibitemOpen
  \bibfield  {author} {\bibinfo {author} {\bibfnamefont {V.}~\bibnamefont
  {Berezinsky}},\ }\bibfield  {title} {\bibinfo {title} {Destruction of
  long-range order in one-dimensional and two-dimensional systems possessing a
  continuous symmetry group. ii. quantum systems.},\ }\href
  {https://inspirehep.net/literature/61186} {\bibfield  {journal} {\bibinfo
  {journal} {Zh. Eksp. Teor. Fiz.}\ }\textbf {\bibinfo {volume} {61}},\
  \bibinfo {pages} {610} (\bibinfo {year} {1972})}\BibitemShut {NoStop}%
\bibitem [{\citenamefont {Kosterlitz}\ and\ \citenamefont
  {Thouless}(1972)}]{JMKosterlitz_1972}%
  \BibitemOpen
  \bibfield  {author} {\bibinfo {author} {\bibfnamefont {J.~M.}\ \bibnamefont
  {Kosterlitz}}\ and\ \bibinfo {author} {\bibfnamefont {D.~J.}\ \bibnamefont
  {Thouless}},\ }\bibfield  {title} {\bibinfo {title} {Long range order and
  metastability in two dimensional solids and superfluids. (application of
  dislocation theory)},\ }\href {https://doi.org/10.1088/0022-3719/5/11/002}
  {\bibfield  {journal} {\bibinfo  {journal} {J. Phys. C}\ }\textbf {\bibinfo
  {volume} {5}},\ \bibinfo {pages} {L124} (\bibinfo {year} {1972})}\BibitemShut
  {NoStop}%
\bibitem [{\citenamefont {Kosterlitz}\ and\ \citenamefont
  {Thouless}(1973)}]{JMKosterlitz_1973}%
  \BibitemOpen
  \bibfield  {author} {\bibinfo {author} {\bibfnamefont {J.~M.}\ \bibnamefont
  {Kosterlitz}}\ and\ \bibinfo {author} {\bibfnamefont {D.~J.}\ \bibnamefont
  {Thouless}},\ }\bibfield  {title} {\bibinfo {title} {Ordering, metastability
  and phase transitions in two-dimensional systems},\ }\href
  {https://doi.org/10.1088/0022-3719/6/7/010} {\bibfield  {journal} {\bibinfo
  {journal} {J. Phys. C}\ }\textbf {\bibinfo {volume} {6}},\ \bibinfo {pages}
  {1181} (\bibinfo {year} {1973})}\BibitemShut {NoStop}%
\bibitem [{\citenamefont {Kosterlitz}(1974)}]{JMKosterlitz_1974}%
  \BibitemOpen
  \bibfield  {author} {\bibinfo {author} {\bibfnamefont {J.~M.}\ \bibnamefont
  {Kosterlitz}},\ }\bibfield  {title} {\bibinfo {title} {The critical
  properties of the two-dimensional xy model},\ }\href
  {https://doi.org/10.1088/0022-3719/7/6/005} {\bibfield  {journal} {\bibinfo
  {journal} {J. Phys. C}\ }\textbf {\bibinfo {volume} {7}},\ \bibinfo {pages}
  {1046} (\bibinfo {year} {1974})}\BibitemShut {NoStop}%
\bibitem [{\citenamefont {Kosterlitz}(2016)}]{Kosterlitz_2016}%
  \BibitemOpen
  \bibfield  {author} {\bibinfo {author} {\bibfnamefont {J.~M.}\ \bibnamefont
  {Kosterlitz}},\ }\bibfield  {title} {\bibinfo {title} {Kosterlitz–thouless
  physics: a review of key issues},\ }\href
  {https://doi.org/10.1088/0034-4885/79/2/026001} {\bibfield  {journal}
  {\bibinfo  {journal} {Rep. Prog. Phys.}\ }\textbf {\bibinfo {volume} {79}},\
  \bibinfo {pages} {026001} (\bibinfo {year} {2016})}\BibitemShut {NoStop}%
\bibitem [{\citenamefont {Kardar}(2007)}]{Kardar2007}%
  \BibitemOpen
  \bibfield  {author} {\bibinfo {author} {\bibfnamefont {M.}~\bibnamefont
  {Kardar}},\ }\href@noop {} {\emph {\bibinfo {title} {Statistical Physics of
  Fields}}}\ (\bibinfo  {publisher} {Cambridge University Press},\ \bibinfo
  {year} {2007})\BibitemShut {NoStop}%
\bibitem [{\citenamefont {Hadzibabic}\ \emph {et~al.}(2006)\citenamefont
  {Hadzibabic}, \citenamefont {Kr{\"u}ger}, \citenamefont {Cheneau},
  \citenamefont {Battelier},\ and\ \citenamefont
  {Dalibard}}]{hadzibabic2006berezinskii}%
  \BibitemOpen
  \bibfield  {author} {\bibinfo {author} {\bibfnamefont {Z.}~\bibnamefont
  {Hadzibabic}}, \bibinfo {author} {\bibfnamefont {P.}~\bibnamefont
  {Kr{\"u}ger}}, \bibinfo {author} {\bibfnamefont {M.}~\bibnamefont {Cheneau}},
  \bibinfo {author} {\bibfnamefont {B.}~\bibnamefont {Battelier}},\ and\
  \bibinfo {author} {\bibfnamefont {J.}~\bibnamefont {Dalibard}},\ }\bibfield
  {title} {\bibinfo {title} {Berezinskii--kosterlitz--thouless crossover in a
  trapped atomic gas},\ }\href
  {https://doi.org/https://doi.org/10.1038/nature04851} {\bibfield  {journal}
  {\bibinfo  {journal} {Nature}\ }\textbf {\bibinfo {volume} {441}},\ \bibinfo
  {pages} {1118} (\bibinfo {year} {2006})}\BibitemShut {NoStop}%
\bibitem [{\citenamefont {Chomaz}\ \emph {et~al.}(2022)\citenamefont {Chomaz},
  \citenamefont {Ferrier-Barbut}, \citenamefont {Ferlaino}, \citenamefont
  {Laburthe-Tolra}, \citenamefont {Lev},\ and\ \citenamefont
  {Pfau}}]{Chomaz_2023}%
  \BibitemOpen
  \bibfield  {author} {\bibinfo {author} {\bibfnamefont {L.}~\bibnamefont
  {Chomaz}}, \bibinfo {author} {\bibfnamefont {I.}~\bibnamefont
  {Ferrier-Barbut}}, \bibinfo {author} {\bibfnamefont {F.}~\bibnamefont
  {Ferlaino}}, \bibinfo {author} {\bibfnamefont {B.}~\bibnamefont
  {Laburthe-Tolra}}, \bibinfo {author} {\bibfnamefont {B.~L.}\ \bibnamefont
  {Lev}},\ and\ \bibinfo {author} {\bibfnamefont {T.}~\bibnamefont {Pfau}},\
  }\bibfield  {title} {\bibinfo {title} {Dipolar physics: a review of
  experiments with magnetic quantum gases},\ }\href
  {https://doi.org/10.1088/1361-6633/aca814} {\bibfield  {journal} {\bibinfo
  {journal} {Rep. Prog. Phys.}\ }\textbf {\bibinfo {volume} {86}},\ \bibinfo
  {pages} {026401} (\bibinfo {year} {2022})}\BibitemShut {NoStop}%
\bibitem [{\citenamefont {Podolsky}\ \emph {et~al.}(2009)\citenamefont
  {Podolsky}, \citenamefont {Chandrasekharan},\ and\ \citenamefont
  {Vishwanath}}]{PhysRevB.80.214513}%
  \BibitemOpen
  \bibfield  {author} {\bibinfo {author} {\bibfnamefont {D.}~\bibnamefont
  {Podolsky}}, \bibinfo {author} {\bibfnamefont {S.}~\bibnamefont
  {Chandrasekharan}},\ and\ \bibinfo {author} {\bibfnamefont {A.}~\bibnamefont
  {Vishwanath}},\ }\bibfield  {title} {\bibinfo {title} {Phase transitions of
  $s=1$ spinor condensates in an optical lattice},\ }\href
  {https://doi.org/10.1103/PhysRevB.80.214513} {\bibfield  {journal} {\bibinfo
  {journal} {Phys. Rev. B}\ }\textbf {\bibinfo {volume} {80}},\ \bibinfo
  {pages} {214513} (\bibinfo {year} {2009})}\BibitemShut {NoStop}%
\bibitem [{\citenamefont {Sch{\"a}fer}\ \emph {et~al.}(2020)\citenamefont
  {Sch{\"a}fer}, \citenamefont {Fukuhara}, \citenamefont {Sugawa},
  \citenamefont {Takasu},\ and\ \citenamefont {Takahashi}}]{schafer2020tools}%
  \BibitemOpen
  \bibfield  {author} {\bibinfo {author} {\bibfnamefont {F.}~\bibnamefont
  {Sch{\"a}fer}}, \bibinfo {author} {\bibfnamefont {T.}~\bibnamefont
  {Fukuhara}}, \bibinfo {author} {\bibfnamefont {S.}~\bibnamefont {Sugawa}},
  \bibinfo {author} {\bibfnamefont {Y.}~\bibnamefont {Takasu}},\ and\ \bibinfo
  {author} {\bibfnamefont {Y.}~\bibnamefont {Takahashi}},\ }\bibfield  {title}
  {\bibinfo {title} {Tools for quantum simulation with ultracold atoms in
  optical lattices},\ }\href
  {https://doi.org/https://doi.org/10.1038/s42254-020-0195-3} {\bibfield
  {journal} {\bibinfo  {journal} {Nat. Rev. Phys.}\ }\textbf {\bibinfo {volume}
  {2}},\ \bibinfo {pages} {411} (\bibinfo {year} {2020})}\BibitemShut {NoStop}%
\bibitem [{\citenamefont {Zheng}\ \emph {et~al.}(2025)\citenamefont {Zheng},
  \citenamefont {Luo}, \citenamefont {Shen}, \citenamefont {He}, \citenamefont
  {Zhu}, \citenamefont {Liu}, \citenamefont {Zhang}, \citenamefont {Sun},
  \citenamefont {Deng}, \citenamefont {Yuan} \emph
  {et~al.}}]{zheng2025counterflow}%
  \BibitemOpen
  \bibfield  {author} {\bibinfo {author} {\bibfnamefont {Y.-G.}\ \bibnamefont
  {Zheng}}, \bibinfo {author} {\bibfnamefont {A.}~\bibnamefont {Luo}}, \bibinfo
  {author} {\bibfnamefont {Y.-C.}\ \bibnamefont {Shen}}, \bibinfo {author}
  {\bibfnamefont {M.-G.}\ \bibnamefont {He}}, \bibinfo {author} {\bibfnamefont
  {Z.-H.}\ \bibnamefont {Zhu}}, \bibinfo {author} {\bibfnamefont
  {Y.}~\bibnamefont {Liu}}, \bibinfo {author} {\bibfnamefont {W.-Y.}\
  \bibnamefont {Zhang}}, \bibinfo {author} {\bibfnamefont {H.}~\bibnamefont
  {Sun}}, \bibinfo {author} {\bibfnamefont {Y.}~\bibnamefont {Deng}}, \bibinfo
  {author} {\bibfnamefont {Z.-S.}\ \bibnamefont {Yuan}}, \emph {et~al.},\
  }\bibfield  {title} {\bibinfo {title} {Counterflow superfluidity in a
  two-component mott insulator},\ }\href@noop {} {\bibfield  {journal}
  {\bibinfo  {journal} {Nature Physics}\ }\textbf {\bibinfo {volume} {21}},\
  \bibinfo {pages} {208} (\bibinfo {year} {2025})}\BibitemShut {NoStop}%
\bibitem [{\citenamefont {Maccari}\ \emph {et~al.}(2023)\citenamefont
  {Maccari}, \citenamefont {Carlstr\"om},\ and\ \citenamefont
  {Babaev}}]{PhysRevB.107.064501}%
  \BibitemOpen
  \bibfield  {author} {\bibinfo {author} {\bibfnamefont {I.}~\bibnamefont
  {Maccari}}, \bibinfo {author} {\bibfnamefont {J.}~\bibnamefont
  {Carlstr\"om}},\ and\ \bibinfo {author} {\bibfnamefont {E.}~\bibnamefont
  {Babaev}},\ }\bibfield  {title} {\bibinfo {title} {Prediction of
  time-reversal-symmetry breaking fermionic quadrupling condensate in twisted
  bilayer graphene},\ }\href {https://doi.org/10.1103/PhysRevB.107.064501}
  {\bibfield  {journal} {\bibinfo  {journal} {Phys. Rev. B}\ }\textbf {\bibinfo
  {volume} {107}},\ \bibinfo {pages} {064501} (\bibinfo {year}
  {2023})}\BibitemShut {NoStop}%
\bibitem [{\citenamefont {Zeng}\ \emph {et~al.}(2024)\citenamefont {Zeng},
  \citenamefont {Hu}, \citenamefont {Hu}, \citenamefont {You},\ and\
  \citenamefont {Wu}}]{zeng2024high}%
  \BibitemOpen
  \bibfield  {author} {\bibinfo {author} {\bibfnamefont {M.}~\bibnamefont
  {Zeng}}, \bibinfo {author} {\bibfnamefont {L.-H.}\ \bibnamefont {Hu}},
  \bibinfo {author} {\bibfnamefont {H.-Y.}\ \bibnamefont {Hu}}, \bibinfo
  {author} {\bibfnamefont {Y.-Z.}\ \bibnamefont {You}},\ and\ \bibinfo {author}
  {\bibfnamefont {C.}~\bibnamefont {Wu}},\ }\bibfield  {title} {\bibinfo
  {title} {High-order time-reversal symmetry breaking normal state},\
  }\href@noop {} {\bibfield  {journal} {\bibinfo  {journal} {Science China
  Physics, Mechanics \& Astronomy}\ }\textbf {\bibinfo {volume} {67}},\
  \bibinfo {pages} {237411} (\bibinfo {year} {2024})}\BibitemShut {NoStop}%
\bibitem [{\citenamefont {Ge}\ \emph {et~al.}(2024)\citenamefont {Ge},
  \citenamefont {Wang}, \citenamefont {Xing}, \citenamefont {Yin},
  \citenamefont {Wang}, \citenamefont {Shen}, \citenamefont {Lei},
  \citenamefont {Wang},\ and\ \citenamefont {Wang}}]{PhysRevX.14.021025}%
  \BibitemOpen
  \bibfield  {author} {\bibinfo {author} {\bibfnamefont {J.}~\bibnamefont
  {Ge}}, \bibinfo {author} {\bibfnamefont {P.}~\bibnamefont {Wang}}, \bibinfo
  {author} {\bibfnamefont {Y.}~\bibnamefont {Xing}}, \bibinfo {author}
  {\bibfnamefont {Q.}~\bibnamefont {Yin}}, \bibinfo {author} {\bibfnamefont
  {A.}~\bibnamefont {Wang}}, \bibinfo {author} {\bibfnamefont {J.}~\bibnamefont
  {Shen}}, \bibinfo {author} {\bibfnamefont {H.}~\bibnamefont {Lei}}, \bibinfo
  {author} {\bibfnamefont {Z.}~\bibnamefont {Wang}},\ and\ \bibinfo {author}
  {\bibfnamefont {J.}~\bibnamefont {Wang}},\ }\bibfield  {title} {\bibinfo
  {title} {Charge-$4e$ and charge-$6e$ flux quantization and higher charge
  superconductivity in kagome superconductor ring devices},\ }\href
  {https://doi.org/10.1103/PhysRevX.14.021025} {\bibfield  {journal} {\bibinfo
  {journal} {Phys. Rev. X}\ }\textbf {\bibinfo {volume} {14}},\ \bibinfo
  {pages} {021025} (\bibinfo {year} {2024})}\BibitemShut {NoStop}%
\bibitem [{\citenamefont {Drouin-Touchette}\ \emph {et~al.}(2022)\citenamefont
  {Drouin-Touchette}, \citenamefont {Orth}, \citenamefont {Coleman},
  \citenamefont {Chandra},\ and\ \citenamefont
  {Lubensky}}]{drouin2022emergent}%
  \BibitemOpen
  \bibfield  {author} {\bibinfo {author} {\bibfnamefont {V.}~\bibnamefont
  {Drouin-Touchette}}, \bibinfo {author} {\bibfnamefont {P.~P.}\ \bibnamefont
  {Orth}}, \bibinfo {author} {\bibfnamefont {P.}~\bibnamefont {Coleman}},
  \bibinfo {author} {\bibfnamefont {P.}~\bibnamefont {Chandra}},\ and\ \bibinfo
  {author} {\bibfnamefont {T.~C.}\ \bibnamefont {Lubensky}},\ }\bibfield
  {title} {\bibinfo {title} {Emergent potts order in a coupled hexatic-nematic
  xy model},\ }\href@noop {} {\bibfield  {journal} {\bibinfo  {journal}
  {Physical Review X}\ }\textbf {\bibinfo {volume} {12}},\ \bibinfo {pages}
  {011043} (\bibinfo {year} {2022})}\BibitemShut {NoStop}%
\bibitem [{\citenamefont {Rydow}\ \emph {et~al.}(2024)\citenamefont {Rydow},
  \citenamefont {Singh}, \citenamefont {Beregi}, \citenamefont {Chang},
  \citenamefont {Mathey}, \citenamefont {Foot},\ and\ \citenamefont
  {Sunami}}]{rydow2024observation}%
  \BibitemOpen
  \bibfield  {author} {\bibinfo {author} {\bibfnamefont {E.}~\bibnamefont
  {Rydow}}, \bibinfo {author} {\bibfnamefont {V.~P.}\ \bibnamefont {Singh}},
  \bibinfo {author} {\bibfnamefont {A.}~\bibnamefont {Beregi}}, \bibinfo
  {author} {\bibfnamefont {E.}~\bibnamefont {Chang}}, \bibinfo {author}
  {\bibfnamefont {L.}~\bibnamefont {Mathey}}, \bibinfo {author} {\bibfnamefont
  {C.~J.}\ \bibnamefont {Foot}},\ and\ \bibinfo {author} {\bibfnamefont
  {S.}~\bibnamefont {Sunami}},\ }\bibfield  {title} {\bibinfo {title}
  {Observation of a bilayer superfluid with interlayer coherence},\ }\href@noop
  {} {\bibfield  {journal} {\bibinfo  {journal} {arXiv preprint
  arXiv:2410.22326}\ } (\bibinfo {year} {2024})}\BibitemShut {NoStop}%
\bibitem [{\citenamefont {Tomita}\ and\ \citenamefont
  {Okabe}(2002)}]{PhysRevB.65.184405}%
  \BibitemOpen
  \bibfield  {author} {\bibinfo {author} {\bibfnamefont {Y.}~\bibnamefont
  {Tomita}}\ and\ \bibinfo {author} {\bibfnamefont {Y.}~\bibnamefont {Okabe}},\
  }\bibfield  {title} {\bibinfo {title} {Probability-changing cluster algorithm
  for two-dimensional $\mathrm{XY}$ and clock models},\ }\href
  {https://doi.org/10.1103/PhysRevB.65.184405} {\bibfield  {journal} {\bibinfo
  {journal} {Phys. Rev. B}\ }\textbf {\bibinfo {volume} {65}},\ \bibinfo
  {pages} {184405} (\bibinfo {year} {2002})}\BibitemShut {NoStop}%
\bibitem [{\citenamefont {Komura}\ and\ \citenamefont
  {Okabe}(2012)}]{komura2012a}%
  \BibitemOpen
  \bibfield  {author} {\bibinfo {author} {\bibfnamefont {Y.}~\bibnamefont
  {Komura}}\ and\ \bibinfo {author} {\bibfnamefont {Y.}~\bibnamefont {Okabe}},\
  }\bibfield  {title} {\bibinfo {title} {Large-{{Scale Monte Carlo Simulation}}
  of {{Two-Dimensional Classical XY Model Using Multiple GPUs}}},\ }\href
  {https://doi.org/10.1143/JPSJ.81.113001} {\bibfield  {journal} {\bibinfo
  {journal} {J. Phys. Soc. Japan.}\ }\textbf {\bibinfo {volume} {81}},\
  \bibinfo {pages} {113001} (\bibinfo {year} {2012})}\BibitemShut {NoStop}%
\bibitem [{\citenamefont {Nienhuis}(1984)}]{nienhuis1984critical}%
  \BibitemOpen
  \bibfield  {author} {\bibinfo {author} {\bibfnamefont {B.}~\bibnamefont
  {Nienhuis}},\ }\bibfield  {title} {\bibinfo {title} {Critical behavior of
  two-dimensional spin models and charge asymmetry in the coulomb gas},\ }\href
  {https://link.springer.com/article/10.1007/BF01009437} {\bibfield  {journal}
  {\bibinfo  {journal} {Stat. Phys}\ }\textbf {\bibinfo {volume} {34}},\
  \bibinfo {pages} {731} (\bibinfo {year} {1984})}\BibitemShut {NoStop}%
\bibitem [{\citenamefont {Swendsen}\ and\ \citenamefont
  {Wang}(1987)}]{PhysRevLett.58.86}%
  \BibitemOpen
  \bibfield  {author} {\bibinfo {author} {\bibfnamefont {R.~H.}\ \bibnamefont
  {Swendsen}}\ and\ \bibinfo {author} {\bibfnamefont {J.-S.}\ \bibnamefont
  {Wang}},\ }\bibfield  {title} {\bibinfo {title} {Nonuniversal critical
  dynamics in monte carlo simulations},\ }\href
  {https://doi.org/10.1103/PhysRevLett.58.86} {\bibfield  {journal} {\bibinfo
  {journal} {Phys. Rev. Lett.}\ }\textbf {\bibinfo {volume} {58}},\ \bibinfo
  {pages} {86} (\bibinfo {year} {1987})}\BibitemShut {NoStop}%
\bibitem [{\citenamefont {Wolff}(1989)}]{PhysRevLett.62.361}%
  \BibitemOpen
  \bibfield  {author} {\bibinfo {author} {\bibfnamefont {U.}~\bibnamefont
  {Wolff}},\ }\bibfield  {title} {\bibinfo {title} {Collective monte carlo
  updating for spin systems},\ }\href
  {https://doi.org/10.1103/PhysRevLett.62.361} {\bibfield  {journal} {\bibinfo
  {journal} {Phys. Rev. Lett.}\ }\textbf {\bibinfo {volume} {62}},\ \bibinfo
  {pages} {361} (\bibinfo {year} {1989})}\BibitemShut {NoStop}%
\bibitem [{\citenamefont {Metropolis}\ \emph {et~al.}(1953)\citenamefont
  {Metropolis}, \citenamefont {Rosenbluth}, \citenamefont {Rosenbluth},
  \citenamefont {Teller},\ and\ \citenamefont {Teller}}]{Metropolis}%
  \BibitemOpen
  \bibfield  {author} {\bibinfo {author} {\bibfnamefont {N.}~\bibnamefont
  {Metropolis}}, \bibinfo {author} {\bibfnamefont {A.~W.}\ \bibnamefont
  {Rosenbluth}}, \bibinfo {author} {\bibfnamefont {M.~N.}\ \bibnamefont
  {Rosenbluth}}, \bibinfo {author} {\bibfnamefont {A.~H.}\ \bibnamefont
  {Teller}},\ and\ \bibinfo {author} {\bibfnamefont {E.}~\bibnamefont
  {Teller}},\ }\bibfield  {title} {\bibinfo {title} {Equation of state
  calculations by fast computing machines},\ }\href
  {https://doi.org/10.1063/1.1699114} {\bibfield  {journal} {\bibinfo
  {journal} {J. Chem. Phys.}\ }\textbf {\bibinfo {volume} {21}},\ \bibinfo
  {pages} {1087} (\bibinfo {year} {1953})}\BibitemShut {NoStop}%
\bibitem [{\citenamefont {Tuan}\ \emph {et~al.}(2022)\citenamefont {Tuan},
  \citenamefont {Long}, \citenamefont {Nui}, \citenamefont {Minh},
  \citenamefont {Trung~Kien},\ and\ \citenamefont
  {Viet}}]{PhysRevE.106.034138}%
  \BibitemOpen
  \bibfield  {author} {\bibinfo {author} {\bibfnamefont {L.~M.}\ \bibnamefont
  {Tuan}}, \bibinfo {author} {\bibfnamefont {T.~T.}\ \bibnamefont {Long}},
  \bibinfo {author} {\bibfnamefont {D.~X.}\ \bibnamefont {Nui}}, \bibinfo
  {author} {\bibfnamefont {P.~T.}\ \bibnamefont {Minh}}, \bibinfo {author}
  {\bibfnamefont {N.~D.}\ \bibnamefont {Trung~Kien}},\ and\ \bibinfo {author}
  {\bibfnamefont {D.~X.}\ \bibnamefont {Viet}},\ }\bibfield  {title} {\bibinfo
  {title} {Binder ratio in the two-dimensional $q$-state clock model},\ }\href
  {https://doi.org/10.1103/PhysRevE.106.034138} {\bibfield  {journal} {\bibinfo
   {journal} {Phys. Rev. E}\ }\textbf {\bibinfo {volume} {106}},\ \bibinfo
  {pages} {034138} (\bibinfo {year} {2022})}\BibitemShut {NoStop}%
\bibitem [{\citenamefont {Viet}\ and\ \citenamefont
  {Kawamura}(2009)}]{PhysRevB.80.064418}%
  \BibitemOpen
  \bibfield  {author} {\bibinfo {author} {\bibfnamefont {D.~X.}\ \bibnamefont
  {Viet}}\ and\ \bibinfo {author} {\bibfnamefont {H.}~\bibnamefont
  {Kawamura}},\ }\bibfield  {title} {\bibinfo {title} {Monte carlo studies of
  chiral and spin ordering of the three-dimensional heisenberg spin glass},\
  }\href {https://doi.org/10.1103/PhysRevB.80.064418} {\bibfield  {journal}
  {\bibinfo  {journal} {Phys. Rev. B}\ }\textbf {\bibinfo {volume} {80}},\
  \bibinfo {pages} {064418} (\bibinfo {year} {2009})}\BibitemShut {NoStop}%
\bibitem [{\citenamefont {Ding}\ \emph {et~al.}(2014)\citenamefont {Ding},
  \citenamefont {Guo},\ and\ \citenamefont {Deng}}]{PhysRevB.90.134420}%
  \BibitemOpen
  \bibfield  {author} {\bibinfo {author} {\bibfnamefont {C.}~\bibnamefont
  {Ding}}, \bibinfo {author} {\bibfnamefont {W.}~\bibnamefont {Guo}},\ and\
  \bibinfo {author} {\bibfnamefont {Y.}~\bibnamefont {Deng}},\ }\bibfield
  {title} {\bibinfo {title} {Reentrance of berezinskii-kosterlitz-thouless-like
  transitions in a three-state potts antiferromagnetic thin film},\ }\href
  {https://doi.org/10.1103/PhysRevB.90.134420} {\bibfield  {journal} {\bibinfo
  {journal} {Phys. Rev. B}\ }\textbf {\bibinfo {volume} {90}},\ \bibinfo
  {pages} {134420} (\bibinfo {year} {2014})}\BibitemShut {NoStop}%
\bibitem [{\citenamefont {Weber}\ and\ \citenamefont
  {Minnhagen}(1988)}]{PhysRevB.37.5986}%
  \BibitemOpen
  \bibfield  {author} {\bibinfo {author} {\bibfnamefont {H.}~\bibnamefont
  {Weber}}\ and\ \bibinfo {author} {\bibfnamefont {P.}~\bibnamefont
  {Minnhagen}},\ }\bibfield  {title} {\bibinfo {title} {Monte carlo
  determination of the critical temperature for the two-dimensional xy model},\
  }\href {https://doi.org/10.1103/PhysRevB.37.5986} {\bibfield  {journal}
  {\bibinfo  {journal} {Phys. Rev. B}\ }\textbf {\bibinfo {volume} {37}},\
  \bibinfo {pages} {5986} (\bibinfo {year} {1988})}\BibitemShut {NoStop}%
\bibitem [{\citenamefont {Janke}(1997)}]{PhysRevB.55.3580}%
  \BibitemOpen
  \bibfield  {author} {\bibinfo {author} {\bibfnamefont {W.}~\bibnamefont
  {Janke}},\ }\bibfield  {title} {\bibinfo {title} {Logarithmic corrections in
  the two-dimensional xy model},\ }\href
  {https://doi.org/10.1103/PhysRevB.55.3580} {\bibfield  {journal} {\bibinfo
  {journal} {Phys. Rev. B}\ }\textbf {\bibinfo {volume} {55}},\ \bibinfo
  {pages} {3580} (\bibinfo {year} {1997})}\BibitemShut {NoStop}%
\bibitem [{\citenamefont {Kenna}\ and\ \citenamefont
  {Irving}(1997)}]{KENNA1997583}%
  \BibitemOpen
  \bibfield  {author} {\bibinfo {author} {\bibfnamefont {R.}~\bibnamefont
  {Kenna}}\ and\ \bibinfo {author} {\bibfnamefont {A.}~\bibnamefont {Irving}},\
  }\bibfield  {title} {\bibinfo {title} {The kosterlitz-thouless universality
  class},\ }\href
  {https://doi.org/https://doi.org/10.1016/S0550-3213(96)00642-6} {\bibfield
  {journal} {\bibinfo  {journal} {Nucl. Phys. B.}\ }\textbf {\bibinfo {volume}
  {485}},\ \bibinfo {pages} {583} (\bibinfo {year} {1997})}\BibitemShut
  {NoStop}%
\bibitem [{ten()}]{tensofermi2025data}%
  \BibitemOpen
  \href@noop {} {\bibinfo {title} {Code and raw data of this paper.}},\
  \bibinfo {howpublished}
  {\url{https://github.com/Tensofermi/Bilayer_XY_Model}}\BibitemShut {NoStop}%
\end{thebibliography}%

\end{document}